\documentclass[showpacs,preprintnumbers,amsmath,amssymb]{revtex4}

\usepackage[dvips]{graphicx}
\usepackage{dcolumn}
\usepackage{bm}

\def\mapgeq{\mathbin{\lower.3ex\hbox{$\buildrel>\over{\smash{\scriptstyle\sim}\vphantom{_x}}$}}}
\def\mapleq{\mathbin{\lower.3ex\hbox{$\buildrel<\over{\smash{\scriptstyle\sim}\vphantom{_x}}$}}}
\def\mapgeqeq{\mathbi{\lower.3ex\hbox{$\buildrel>\over{\smash{\scriptstyle\approx}\vphantom{_2}}$}}}
\def\mapleqeq{\mathbin{\lower.3ex\hbox{$\buildrel<\over{\smash{\scriptstyle\approx}\vphantom{_2}}$}}}
\def\Journal#1#2#3#4{{#1} {\bf #2} (#4) #3}
\def\MPL{Mod. Phys. Lett. A}

\def\NPB{Nucl. Phys. B}
\def\NPBOLD{Nucl. Phys.}

\def\PLB{{Phys. Lett.} B}

\def\PLBOLD{Phys. Lett.}
\def\PRL{Phys. Rev. Lett.}

\def\PRD{Phys. Rev. D}

\def\PTP{Prog. Theor. Phys.}
\def\JHEP{JHEP}
\def\EPJ{Euro. Phys. J. C}
\def\JETPUSSR{JETP (USSR)}
\def\ZETP{Zh. Eksp. Teor. Piz.}

\def\IJMP{Int. J. Mod. Phys. A}

\begin{document}

\preprint{TOKAI-HEP/TH-0202}

\title{$S_{2L}$ permutation symmetry for left-handed $\mu$ and $\tau$ families \\
       and neutrino oscillations in an $SU(3)_L \times U(1)_N$ gauge model}

\author{Teruyuki Kitabayashi}
 \email{teruyuki@post.kek.jp}
\affiliation{%
\sl Accelerator Engineering Center, Mitsubishi Electric System \& Service Engineering Co.Ltd.,
\\
2-8-8 Umezono, Tsukuba, Ibaraki 305-0045, Japan
}%

\author{Masaki Yasu\`{e}}
\email{yasue@keyaki.cc.u-tokai.ac.jp}
\affiliation{%
\sl General Education Program Center, Shimizu Campus,\\
School of Marine Science and Technology, Tokai University,\\
3-20-1 Orido, Shimizu, Shizuoka 424-8610, Japan\\
{\rm and}\\
Department of Physics, Tokai University,\\
1117 Kitakaname, Hiratsuka,
\\Kanagawa 259-1291, Japan.}%

\date{September, 2002}

\begin{abstract}
We construct an $SU(3)_L \times U(1)_N$ gauge model based on an $S_{2L}$ permutation symmetry for left-handed $\mu$ and $\tau$ families, which provides the almost maximal atmospheric neutrino mixing and the large solar neutrino mixing of the LMA type. Neutrinos acquire one-loop radiative masses induced by the radiative mechanism of the Zee type as well as tree level masses induced by the type II seesaw mechanism utilizing interactions of lepton triplets with an $SU(3)$-sextet scalar. The atmospheric neutrino mixing controlled by the tree-level and radiative masses turns out to be almost maximal owing to the presence of $S_{2L}$ supplemented by a $Z_4$ discrete symmetry. These symmetries ensure the almost equality between the $\nu_e$-$\nu_\mu$ and $\nu_e$-$\nu_\tau$ radiative masses dominated by contributions from heavy leptons contained in the third members of lepton triplets, whose Yukawa interactions conserve $S_{2L}$ even after the spontaneous breaking. The solar neutrino mixing controlled by radiative masses including a $\nu_\mu$-$\nu_\tau$ mass, which are taken to be the similar order, turns out to be described by large solar neutrino mixing angles.
\end{abstract}

\maketitle

Recent observations of atmospheric and solar neutrino oscillations have provided the clear evidence that neutrinos are massive particles \cite{Kamiokande, SNO, SNOexpRelated}. These oscillations are characterized by the squared mass differences for atmospheric neutrinos, $\Delta m_{atm}^2$ $\sim$ $3\times10^{-3}$ eV$^2$, with the mixing angle of $\sin^2 2\theta_{atm} \sim 1$ \cite{RecentAtm} and for solar neutrinos, $\Delta m_\odot^2 \sim 10^{-5}-10^{-4}$ eV$^2$, for the LMA solution with $\sin^2 2\theta_\odot \sim 0.75$ and, $\Delta m_\odot^2 \sim 10^{-8}$ eV$^2$, for the LOW solution with $\sin^2 2\theta_\odot \sim 0.92$ \cite{SNOexpRelated, RecentSolar}, where the LMA solution is currently considered as the most favorable solution. The existence of these massive neutrinos and their oscillations requires some new interactions beyond the conventional interactions in the standard model \cite{MassiveNeutrino}. Furthermore, the data indicate a mass hierarchy of $\Delta m_{atm}^2$ $\gg$ $\Delta m_\odot^2$ as well as the large mixing angles, suggesting that the neutrino mass matrix has bimaximal structure \cite{Bimaximal, NearlyBimaximal}. 

There are two main theoretical mechanisms to generate tiny neutrino masses: one is the seesaw mechanism \cite{Seesaw, type2seesaw} and the other is the radiative mechanism \cite{Zee, Babu}. It has been pointed out that the radiative mechanism of the Zee type \cite{Zee} may fail to explain the favorable LMA solution \cite{ZeeMaximal}. In the Zee model, a Higgs scalar $\phi^\prime$ as a duplicate of the standard Higgs scalar $\phi$ and a singly charged $SU(2)_L$-singlet scalar $h^+$ have been introduced into the standard model to generate tiny neutrino masses by one-loop radiative corrections. The neutrino mass matrix in the Zee model has the following form:

\begin{equation}
M_\nu^{rad} =    
    \left(
    \begin{array}{ccc}
         0       & b^{rad} & c^{rad} \\
         b^{rad} & 0       & e^{rad} \\
         c^{rad} & e^{rad} & 0\\
    \end{array}
    \right),
	\label{Eq:Mrad}
\end{equation}
where $b^{rad}, c^{rad}$ and $e^{rad}$ stand for radiatively induced neutrino masses. There is an obvious relation among the neutrino masses denoted by $m_{1,2,3}$, dictating $m_1+m_2+m_3=0$, from which the mixing angle for solar neutrinos is constrained to $\sin^2 2\theta_\odot \sim 1.0$ \cite{ZeeMaximal}. However, recent observations show that the best-fit value of the mixing angle for the LMA solution is $\sin^2 2\theta_\odot \sim 0.8$. Thus, the original Zee model is not capable of explaining neutrino oscillations compatible with the LMA solution.

 To implement the radiative mechanism of the Zee type, we have advocated to use a triplet Higgs scalar in $SU(3)_L \times U(1)_N$ gauge models \cite{331}. The Zee scalar $h^+$ is identified with the third member of an $SU(3)_L$-triplet Higgs scalar and can be unified into a triplet $\eta$ with the standard Higgs doublet $(\phi^0,\phi^-)$. Namely, an $SU(3)_L$-triplet $(\eta^0, \eta^-, \eta^+)$ can be interpreted as $(\phi^0,\phi^-,h^+)$. Therefore, the existence of Zee scalar $h^+$ is naturally understood. Furthermore, the $SU(3)_L \times U(1)_N$ gauge models are known to exhibit the attractive properties that these models predict three families of quarks and leptons from the anomaly free conditions on $SU(3)_L \times U(1)_N$ and the asymptotic freedom of $SU(3)_c$. The anomalies are cancelled by the six triplets and six antitriplets, which are appropriately supplied by three families of leptons and three families of three colors of quarks. It is remarkable that this cancellation mechanism only works in the multiple of three families. With this plausible properties, radiative mechanisms to generate tiny masses of neutrinos and their oscillations have been extensively studied in $SU(3)_L \times U(1)_N$ gauge models \cite{331LeptonMasses, Zee331}. However, the possibility of explaining the observed properties of solar neutrino oscillations with $\sin^2 2\theta_\odot \sim 0.8$ has not been emphasized yet.

In this paper, we consider a radiative mechanism of the Zee type in the $SU(3)_L \times U(1)_N$ framework \cite{331} to explain observed properties of neutrino oscillations consistent with the LMA solution of $\sin^2 2\theta_\odot \sim 0.8$. To accommodate the LMA solution to the radiative mechanism of the Zee type, we have to add some ingredients to the model in order to avoid the constraint of $m_1+m_2+m_3=0$. For example, we can obtain the desirable LMA solution (1) if we allow the duplicate Higgs scalar $\phi^\prime$, which is constrained to couple to no leptons in the original Zee model, to couple to leptons \cite{ZeeLMA1}, (2) if we implement an $S_2$ permutation symmetry for the $\mu$ and $\tau$ families with a triplet Higgs scalar \cite{ZeeLMA2}, (3) if we import a sterile neutrino into the model \cite{ZeeLMA3}, and (4) if we make use of (anti) sleptons in supersymmetric gauge models \cite{ZeeLMA4}. In this paper, we examine the phenomena of neutrino oscillations based on the case (2). We show that 

\begin{enumerate}
\item The almost maximal atmospheric neutrino mixing is ensured by the presence of an $S_{2L}$ permutation symmetry for the left-handed states in the $\mu$ and $\tau$ families supplemented by a $Z_4$ discrete symmetry. Especially, in order to explain the maximal mixing, we do not need fine-tuning of couplings of leptons \cite{inverseHierarchy} to the Zee scalar $h^+$, which is now contained in the third member of a triplet Higgs scalar. Instead, heavy leptons take care of the relevant couplings, which dynamically assure the appearance of the maximal mixing.

\item The large solar neutrino mixing characterized by $\sin^2 2\theta_\odot \sim 0.8$ is realized to occur if the magnitudes of radiatively induced neutrino masses are kept to be the same order.
\end{enumerate}

In the next section, we present a possible texture of the neutrino mass matrix with $S_{2L}$ and $Z_4$ symmetries to lead the observed properties of the neutrino mixings and discuss which elements affect the patterns of these mixings. In Sec.\ref{sec:3}, we construct an $SU(3)_L \times U(1)_N$ gauge model to accommodate the observed atmospheric and solar neutrino oscillations. How to generate the neutrino masses and mixings in our model are discussed in Sec.\ref{sec:4}. The results of our detailed analysis are discussed in Sec.\ref{sec:5}. The final section is devoted to summary.

\section{\label{sec:2}Texture of the neutrino mass matrix and $S_{2L}$ permutation symmetry}
In this section, we discuss the usefulness of the $S_{2L}$ permutation symmetry for $\mu$ and $\tau$ families. Following the expressions used in Ref.\cite{ZeeLMA2}, we parameterize the neutrino mass matrix, $M_\nu$, for $\sin\theta_{13}=0$ with $\sin\theta_{13}$ being the mixing angle between $\nu_e$ and $\nu_\tau$, to be:

\begin{eqnarray}
M_\nu =
    \left(
    \begin{array}{ccc}
         a & b & c(=-\sigma b) \\
         b & d & e \\
         c & e & f(=d+(\sigma^{-1}-\sigma )e) \\
    \end{array}
    \right),
	\label{Eq:Mnu}
\end{eqnarray}
where $\sigma=\tan \theta_{23}$ with $\theta_{23}$ being the mixing angle between $\nu_\mu$ and $\nu_\tau$. Similar textures of the neutrino mass matrix have also been studied in literatures \cite{RecentNeutrinoMassTexture, TextureZeros, MassMatrix}. This mass matrix can be diagonalized by $U_{MNS}$ defined by

\begin{eqnarray}
U_{MNS}=\left( 
    \begin{array}{ccc}
        \cos\theta_{12}                &  \sin\theta_{12}                & 0\\
       -\cos\theta_{23}\sin\theta_{12} &  \cos\theta_{23}\cos\theta_{12} & \sin\theta_{23}\\
        \sin\theta_{23}\sin\theta_{12} & -\sin\theta_{23}\cos\theta_{12} & \cos\theta_{23}\\
    \end{array} 
	\right),
	\label{Eq:Umns}
\end{eqnarray}
where $\theta_{12}=\theta_\odot$ and $\theta_{23}=\theta_{atm}$, which transforms $\vert \nu_{mass} \rangle$ as the mass eigenstate with $(m_1, m_2, m_3)$ into $\vert \nu_{weak} \rangle$ as the weak eigenstate by $\vert \nu_{weak} \rangle = U_{MNS} \vert \nu_{mass} \rangle$. The neutrino masses and mixing angles can be parameterized to be:

\begin{eqnarray}
m_1 &=& a -\frac{1}{2}\sqrt{\frac{b^2+c^2}{2}}\left( x + \sqrt{x^2+8} \right), 
\nonumber \\
m_2 &=& a -\frac{1}{2}\sqrt{\frac{b^2+c^2}{2}}\left( x - \sqrt{x^2+8} \right), 
\nonumber \\
m_3 &=& d + \sigma^{-2}\left( d-a+x\sqrt{\frac{b^2+c^2}{2}} \right), 
\nonumber \\
\sin^2 2\theta_\odot&=&\frac{8}{8+x^2}, \quad
\tan \theta_{atm} = -\frac{b}{c} (\equiv \sigma)
\label{Eq:m_sin}
\end{eqnarray}
with

\begin{eqnarray}
x=\sqrt{\frac{2(a-d+\sigma e)^2}{b^2+c^2}}.
\label{Eq:x2}
\end{eqnarray}
From Eq.(\ref{Eq:m_sin}) and Eq.(\ref{Eq:x2}), we find that the condition to obtain the large mixing angle as the LMA solution such as $\sin^2 2\theta_\odot \sim 0.8$ is $x^2 = \mathcal{O}(1)$ or equivalently 

\begin{eqnarray}
(a-d+\sigma e)^2 = \mathcal{O}(b^2+c^2).
\label{Eq:conditionOfsin08}
\end{eqnarray}
There are various solutions to Eq.(\ref{Eq:conditionOfsin08}). We adopt the solution saturated by radiative neutrino masses of the same order. Since both sides must be the same order, the relation of Eq.(\ref{Eq:conditionOfsin08}) requires the cancellation of tree-level neutrino masses if exist.

We divide the neutrino mass matrix into two parts: (1) a tree level mass matrix $M_\nu^{tree}$ and (2) a radiatively induced mass matrix $M_\nu^{rad}$ as follows:

\begin{eqnarray}
M_\nu &=& M_\nu^{tree} + M_\nu^{rad},
\nonumber \\
M_\nu^{tree} &=&
	\left(
    \begin{array}{ccc}
         0 & 0               & 0 \\
         0 & d^{tree}        & \sigma d^{tree} \\
         0 & \sigma d^{tree} & d^{tree} \\
    \end{array}
    \right),
\quad
M_\nu^{rad} =
	\left(
    \begin{array}{ccc}
         a^{rad} & b^{rad} & c^{rad} \\
         b^{rad} & d^{rad} & e^{rad}\\
         c^{rad} & e^{rad} & f^{rad} \\
    \end{array}
    \right),
\label{Eq:MnuMtreeMrad}
\end{eqnarray}
where the superscripts, ``tree'' and ``rad'', denote the tree-level and radiative masses, respectively. The form of $M_\nu^{tree}$ is to be ensured by introducing the $S_{2L}$ permutation symmetry for left-handed states in the $\mu$ and $\tau$ families. An additional $Z_4$ discrete symmetry will pick up the solution with either $\sigma=1$ or $\sigma=-1$. The form of $M_\nu^{tree}$ with $\sigma=\pm 1$ leads to the cancellation of the tree-level masses in $a-d+\sigma e$ and Eq.(\ref{Eq:conditionOfsin08}) required for $\sin^2 2\theta_\odot \sim 0.8$ is transformed to

\begin{eqnarray}
(a^{rad}-d^{rad}+\sigma e^{rad})^2=\mathcal{O}( (b^{rad})^2 + (c^{rad})^2).
\label{Eq:conditionOfsin08_no2}
\end{eqnarray}
If the magnitude of these neutrino masses is kept almost the same to satisfy Eq.(\ref{Eq:conditionOfsin08_no2}), we obtain the significant deviation of $\sin^2 2\theta_\odot$ from unity.   More precisely, at least one of $e-e$ and $\mu-\tau$ sectors provides the same order of magnitude as the $e-\mu$ and $e-\tau$ masses.

So far, these arguments are entirely based on the relation of $f=d+(\sigma^{-1}-\sigma )e$ in Eq.(\ref{Eq:Mnu}). However, since radiatively generated masses may also randomly contribute in $d$, $e$ and $f$, these contributions jeopardize the relation. The effects from the radiative masses cause $\sin \theta_{13} \neq 0$, leading to $U_{e3} \neq 0$, and can be estimated by the conventional perturbative treatment because these effects are much smaller than those from the tree level masses. Denoting the deviation by $\epsilon=f-[d+(\sigma^{-1}-\sigma )e]$, we parameterize $M_\nu$ as:

\begin{eqnarray}
   M_\nu=
	\left(
    \begin{array}{ccc}
         a         & b & -\sigma b \\
         b         & d & e \\
         -\sigma b & e & d+(\sigma^{-1}-\sigma )e \\
    \end{array}
    \right)
	+
	\left(
    \begin{array}{ccc}
         0 & 0 & 0 \\
         0 & 0 & 0 \\
         0 & 0 & \epsilon \\
    \end{array}
    \right).
	\label{Eq:MnuForUe3}
\end{eqnarray}
We find that the lepton mixing angles of $\theta_{12}$ and $\theta_{23}$ are modified into $\theta_{12}^{obs}=\theta_{12}+\xi_{12}$ and $\theta_{23}^{obs}=\theta_{23}+\xi_{23}$, where 
\begin{eqnarray}
\xi_{12}=\frac{s_{23}^2 c_{12} s_{12}}{m_2-m_1}\epsilon, \quad
\xi_{23}= c_{23} s_{23} \left (\frac{s_{12}^2}{m_3-m_1}+\frac{c_{12}^2}{m_3-m_2} \right)\epsilon,
\label{Eq:xi12_xi23}
\end{eqnarray}
and

\begin{eqnarray}
U_{e3}=c_{23} s_{23} c_{12} s_{12} \left (\frac{1}{m_3-m_2}-\frac{1}{m_3-m_1} \right) \epsilon \sim c_{23} s_{23} c_{12} s_{12}\frac{m_2-m_1}{m_3} \frac{\epsilon}{m_3},
\label{Eq:Ue3}
\end{eqnarray}
where $m_3\gg m_{1,2}$ is applied. Furthermore, since we are anticipating that $m_2 \pm m_1 \sim \delta m_\nu^{rad}$ representing radiatively generated neutrino masses, we roughly obtain $m_2-m_1\sim \sqrt{\Delta m_\odot^2}$ and similarly $m_3\sim \sqrt{\Delta m_{atm}^2}$, Eq.({\ref{Eq:Ue3}) can be reduced to

\begin{eqnarray}
\vert U_{e3}\vert\sim \bigg|\frac{\epsilon\sqrt{\Delta m_\odot^2}}{\Delta m_{atm}^2}\bigg| \sim {\rm a~few}\times \frac{\epsilon}{\rm 1~eV},
\label{Eq:Ue3Order}
\end{eqnarray}
where we have used $c_{12}\sim s_{12}\sim c_{23} \sim s_{23} \sim 1$.  Experimentally, the CHOOZ and PALOVERDE data imply that $U_{e3}$ is close to zero, e.g., $\vert U_{e3} \vert^2 \le 0.015-0.05$ \cite{Ue3}, which should be satisfied by Eq.(\ref{Eq:Ue3}).

\section{\label{sec:3}Model}
We choose the $SU(3)_L \times U(1)_N$ gauge model employed in Ref.\cite{ref331} as the reference model, where the leptons are assigned to be:

\begin{eqnarray}
\psi^i_L = \left( \nu^i,\ell^i,\kappa^i\right)_L^T : \left( \textbf{3}, 0 \right), \quad
\ell^{e,\mu,\tau}_R : \left( \textbf{1}, -1 \right),  \quad  
\kappa^{e,-,+}_R    : \left( \textbf{1}, 1 \right).
\label{Eq:leptons}
\end{eqnarray}
Here, the index of $i$=($e,\mu,\tau$) denotes the three families and $\kappa_R^j$ for $j$ = ($e,-,+$) are the mass eigenstates of the positively charged heavy leptons. The superscripts, $\pm$, of $\kappa_R$ correspond to the $\tau \pm \mu$ states of $\kappa_L$ as the chiral partners to be defined by their Yukawa interactions. Higgs scalars are assigned to be:
\begin{eqnarray}
\eta = \left(\eta^0,\eta^-,\eta^+    \right)^T : \left( \textbf{3}, 0 \right), \quad
\rho = \left(\rho^+,\rho^0,\rho^{++} \right)^T : \left( \textbf{3}, 1 \right), \quad
\chi = \left(\chi^-,\chi^{--},\chi^0 \right)^T : \left( \textbf{3}, -1 \right).
\label{Eq:higgs}
\end{eqnarray}
The quantum numbers are specified in parentheses by $(SU(3)_L, U(1)_N)$. Let $N/2$ be the $U(1)_N$ number, then the hypercharge $Y$ and the electric charge $Q$ are given by $Y=-\sqrt{3}\lambda^8+N$ and $Q=(\lambda^3+Y)/2$ respectively, where $\lambda^a$ are the Gell-Mann matrices with Tr$(\lambda^a \lambda^b)=2\delta^{ab} (a,b=1,2,...,8)$. The Higgs scalars develop the following vacuum expectation values (VEV's):

\begin{eqnarray}
\langle 0 \vert\eta\vert 0 \rangle = \left(v_\eta,0,      0      \right)^T, \quad
\langle 0 \vert\rho\vert 0 \rangle = \left(0,     v_\rho, 0      \right)^T, \quad
\langle 0 \vert\chi\vert 0 \rangle = \left(0,     0,      v_\chi \right)^T,
\label{Eq:VEVforEtaRhoChi}
\end{eqnarray}
where the orthogonal choice of these VEV's will be guaranteed by the $\eta\rho\chi$ type Higgs interactions introduced in Eq.(\ref{Eq:higgsV}).

Now, we extend the reference model to our present $SU(3)_L \times U(1)_N$ model in order to accommodate the maximal atmospheric and large solar neutrino mixings. We introduce an $S_{2L}$ permutation symmetry for left-handed states in the $\mu$ and $\tau$ families used together with an additional $Z_4$ discrete symmetry. The quantum numbers of $S_{2L}$ and $Z_4$ as well as the lepton ($L$) and the electron ($L_e$) numbers are listed in Table \ref{Tab:particlesAndSymmetries} for all participating particles in our discussions, where $\psi_L^\pm=(\psi_L^\tau \pm \psi_L^\mu)/\sqrt{2}$. To generate the tree level neutrino masses, we introduce an $S_{2L}$-symmetric anti-sextet scalar, $s$, defined by

\begin{eqnarray}
s= \left(
    \begin{array}{ccc}
         s_\nu^0 & s^+      & s^-      \\
         s^+     & s^{++}   & s_\ell^0 \\
         s^-     & s_\ell^0 & s^{--}   \\
    \end{array}
    \right): \left( \textbf{6}^\ast, 0 \right).
\label{Eq:s}
\end{eqnarray}
Then, the $S_2$- and $Z_4$-conserved interaction of

\begin{eqnarray}
g_s^+ \overline{(\psi_{L\alpha}^-)^c}s^{\alpha \beta}\psi_{L\beta}^-,
\label{Eq:treeMassInt}
\end{eqnarray}
accounts for the form of $M_\nu^{tree}$ with $\sigma=-1$, where $\alpha,\beta$ and $\gamma$ denote the $SU(3)_L$ indices.  We also introduce $S_{2L}$-antisymmetric scalars 

\begin{eqnarray}
\eta^\prime=(\eta^{\prime 0},\eta^{\prime -},\eta^{\prime +})^T:(\textbf{3},0), \quad
\rho^\prime=(\rho^{\prime +},\rho^{\prime 0},\rho^{\prime ++})^T : (\textbf{3},1)
\label{Eq:etaPimeRhoPrime}
\end{eqnarray}
with VEV's of 

\begin{eqnarray}
\langle 0 \vert \eta^\prime \vert 0 \rangle = (v_{\eta^\prime},0,0)^T, \quad
\langle 0 \vert \rho^\prime \vert 0 \rangle = (0,v_{\rho^\prime},0)^T.
\label{Eq:VEVforEtaPrimeRhoPrime}
\end{eqnarray}
These VEV's are also determined by the appropriate Higgs interactions. The scalar $\eta^\prime$ allows us to realize the radiatively induced $\nu_\mu$-$\nu_\tau$ masses by the interaction of 

\begin{eqnarray}
f_{\mu\tau} \epsilon^{\alpha\beta\gamma} \overline{(\psi_{L\alpha}^+)^c}\eta_\beta^\prime \psi_{L\gamma}^-,
\label{Eq:sourceOfNumu_Nutau}
\end{eqnarray}
and the scalar $\rho^\prime$ allows us to realize the mass hierarchy of $m_\mu \ll m_\tau$ as we show later. 

It should be noted that all of these interactions respect the $L_e$ conservation. Furthermore, it is not spoiled by the spontaneous breaking due to VEV's of Higgs scalars as can be seen from Table \ref{Tab:particlesAndSymmetries}. This $L_e$ conservation is, of course, broken by the presence of the $\nu_e$-$\nu_\mu$ and $\nu_e$-$\nu_\tau$ masses, which are radiatively induced by the interactions of

\begin{eqnarray}
f_{e\ell}\epsilon^{\alpha\beta\gamma} \overline{\left( \psi_{L\alpha}^e \right)^c} \eta_\beta \psi_{L\gamma}^+.
\label{Eq:LeViolatingInt}
\end{eqnarray}
As a result, $L_e$ is conserved in all $S_2$- and $Z_4$-invariant Yukawa interactions except $f_{e\ell}\overline{\left( \psi_L^e \right)^c} \eta \psi_L^+$, which can be read off from Table \ref{Tab:S2Z4LLeforYukawa}, where the $S_{2L}$, $Z_4$, $L$ and $L_e$ numbers of the possible Yukawa interactions are listed. This $L_e$-violating interaction can be much suppressed because the limit of $f_{e\ell} \rightarrow 0$ enhances the symmetry of the theory through the restoration of $L_e$ \cite{tHooft}. Another useful symmetry based on $L^\prime (=2L_e-L=L_e - L_\mu - L_\tau)$ \cite{Lprime}, respected by all Yukawa interactions including $\overline{(\psi_L^e)^c}\eta \psi_L^+$, is to be spontaneously broken. But, it is explicitly broken by Higgs interactions such as $\eta s \eta$ in Eq.(\ref{Eq:higgsV}) (see Table \ref{Tab:S2Z4LLeforHiggsInteractions}). So, there is no harmful Nambu-Goldstone boson. In the present discussions, we do not resort to this $L^\prime$ symmetry because the $L_e$- and $S_{2L}$-conservations supersede the $L^\prime$-conservation.

The Yukawa interactions for leptons are now caused by $\mathcal{L}_Y$:

\begin{eqnarray}
-\mathcal{L}_Y &=& 
     f_{e\ell}\overline{\left( \psi_L^e \right)^c} \eta \psi_L^+
   + f_{\mu\tau}\overline{\left( \psi_L^+ \right)^c} \eta^\prime \psi_L^-
   + g_s^+ \overline{\left( \psi_L^- \right)^c} s \psi_L^-
\nonumber \\
&& + f_e \overline{\psi_L^e}\rho e_R
   + \overline{\psi_L^+}\rho        (f_\mu^+ \mu_R + f_\tau^+ \tau_R)
   + \overline{\psi_L^-}\rho^\prime (g_\mu^- \mu_R + g_\tau^- \tau_R)
\nonumber \\
&& + f_\kappa^e \overline{\psi_L^e}\chi \kappa_R^e
   + f_\kappa^+ \overline{\psi_L^+}\chi \kappa_R^+
   + g_\kappa^+ \overline{\psi_L^-}\chi \kappa_R^-
   + (h.c.),
\label{Eq:Yukawa}
\end{eqnarray}
where $f$'s and $g$'s denote the Yukawa couplings. The Higgs interactions are given by Hermitian terms composed of $\phi_\alpha^\dagger \phi_\beta$ $(\phi=\eta, \eta^\prime, \rho, \rho^\prime, \chi, s)$ and by non-Hermitian terms in 

\begin{eqnarray}
V  &=& \mu_1 \eta s \eta
    + \mu_2 \eta^\prime s \eta^\prime
\nonumber \\
    &+& \lambda_1 \epsilon^{\alpha\beta\gamma} \eta_\alpha \rho_\beta \chi_\gamma
    + \lambda_2 \epsilon_{\alpha\beta\gamma} (\eta s)^\alpha \rho^{\dagger_\beta} \chi^{\dagger\gamma} 
    + \lambda_3 (\eta^\dagger\rho^\prime)(\eta^{\prime\dagger}\chi)
    + \lambda_4 (\eta^\dagger\chi)(\eta^{\prime\dagger}\rho^\prime)
    + (h.c.),
\label{Eq:higgsV}
\end{eqnarray}
where $\mu$'s and $\lambda$'s denote a mass scale and coupling constants, respectively. We note that 

\begin{itemize}
\item  the $\eta s \eta$ and $\eta^\prime s \eta^\prime$ terms are the source of the type II seesaw mechanism \cite{type2seesaw}, which calls for the mass of $s$ much greater than $v_{weak}$, the weak scale of $\mathcal{O}(100)$ GeV,

\item The $s^0_\ell$-term of Eq.(\ref{Eq:s}) would induce the dangerous mass-mixings between charged leptons and heavy leptons if $\langle 0 \vert s^0_\ell\vert 0 \rangle \neq 0$. This VEV will be dynamically generated if the potential includes terms such as $(\rho s)\rho^\dagger \eta^\dagger$ and $(\chi s)\chi^\dagger \eta^\dagger$, effectively corresponding to tadpole interactions  of $s^0_\ell$ once VEV's of $\eta$, $\rho$ and $\chi$ are generated.  However, our dynamics regulated by the present potential allows us to set this VEV to vanish.  So, there are no such dangerous mixings.

\item the $\eta\rho\chi$ term ensures the orthogonal choice of VEV's of $\eta,\rho$ and $\chi$,

\item the $\vert \eta^\dagger \eta^\prime \vert^2$- and $\vert \rho^\dagger \rho^\prime \vert^2$-type Higgs interactions present in the Hermitian terms can induce the correct vacuum alignment of Eq.(\ref{Eq:VEVforEtaPrimeRhoPrime}) if their coefficients are taken to be negative.
\end{itemize}
These Higgs interactions are invariant under $S_{2L}$ with $Z_4$ as well as $L_e$ and other interactions are forbidden by these conservations as shown in Table \ref{Tab:S2Z4LLeforHiggsInteractions}. Especially, the absence of $\eta \eta s^c s^c$ is important. This term could yield a divergent mass term of $\nu_e$-$\nu_e$ at the two-loop level as shown in FIG.\ref{Fig:DivergentTwoLoop}; therefore, the tree level mass term is required as a counter term at the $(1,1)$ entry in the $M_\nu^{tree}$ to cancel the divergent. However, this counter term spoils the realization of the texture of $M_\nu^{tree}$. The requirements from $S_{2L}$ and $Z_4$ ensure the internal consistency between the assumed form of $M_\nu^{tree}$ and the absence of this radiative graph.

In the present article, we do not discuss phenomenology due to the existence of heavy leptons and extra gauge bosons as well as heavy exotic quarks \cite{331Related}. Since the standard model well describes the current physics, their contributions should be suppressed. Their masses are controlled by the VEV of $\chi$, which is taken to be $\mathcal{O}(1)$ TeV for the later analyses so that these additional contributions are sufficiently suppressed.

Before discussing how to describe atmospheric and solar neutrino oscillations in our model, we examine the form of mass matrices of the heavy leptons and charged leptons given by Eq.(\ref{Eq:Yukawa}). The heavy lepton mass matrix is simply given by the diagonal masses computed to be $m_\kappa^e=f_\kappa^e v_\chi$, $m_\kappa^-=g_\kappa^+ v_\chi$ and $m_\kappa^+=f_\kappa^+ v_\chi$. On the other hand, the charged lepton mass matrix has the following non-diagonal form:
\begin{eqnarray}
M_\ell= \left(
    \begin{array}{ccc}
    m_\ell^{ee} & 0                & 0                 \\
    0           & m_\ell^{\mu\mu}  & m_\ell^{\mu\tau}  \\
    0           & m_\ell^{\tau\mu} & m_\ell^{\tau\tau} \\
    \end{array}
    \right),
\label{Eq:Mell}
\end{eqnarray}
where
\begin{eqnarray}
m_\ell^{ee}&=&f_ev_\rho,
\nonumber \\
m_\ell^{\mu\mu}&=&\frac{1}{\sqrt{2}} 
    (f_\mu^+ v_\rho - g_\mu^- v_{\rho^\prime}), \quad
m_\ell^{\mu\tau}=\frac{1}{\sqrt{2}} 
    (f_\tau^+ v_\rho - g_\tau^- v_{\rho^\prime}),
\nonumber \\
m_\ell^{\tau\mu}&=&\frac{1}{\sqrt{2}} 
    (f_\mu^+ v_\rho + g_\mu^- v_{\rho^\prime}), \quad
m_\ell^{\tau\tau}=\frac{1}{\sqrt{2}} 
    (f_\tau^+ v_\rho + g_\tau^- v_{\rho^\prime}).
\label{Eq:MellElements}
\end{eqnarray}
The diagonal masses are obtained after the transformation of $M_\ell$ as $M_\ell^{diag} = diag.(m_e,m_\mu,m_\tau)= U_\ell^\dagger M_\ell V_\ell$, where unitary matrices $U_\ell$ and $V_\ell$ are given by 
\begin{eqnarray}
U_\ell = \left(
    \begin{array}{ccc}
    1 & 0 &  0 \\
    0 & c_\alpha & s_\alpha \\
    0 & -s_\alpha & c_\alpha \\
    \end{array}
    \right), 
\quad
V_\ell = \left(
    \begin{array}{ccc}
    1 & 0 &  0 \\
    0 & c_\beta & s_\beta \\
    0 & -s_\beta & c_\beta \\
    \end{array}
    \right) 
\label{Eq:UellVell}
\end{eqnarray}
with $c_\alpha = \cos \alpha,$ etc., defined by
\begin{eqnarray}
c_\alpha &=& \sqrt{\frac{(m_\ell^{\tau\tau})^2+(m_\ell^{\tau\mu})^2-m_\mu^2}{m_\tau^2-m_\mu^2}},
\quad
s_\alpha = \sqrt{\frac{-(m_\ell^{\tau\tau})^2-(m_\ell^{\tau\mu})^2+m_\tau^2}{m_\tau^2-m_\mu^2}},
\nonumber \\
c_\beta &=& \sqrt{\frac{(m_\ell^{\tau\tau})^2+(m_\ell^{\mu\tau})^2-m_\mu^2}{m_\tau^2-m_\mu^2}},
\quad
s_\beta = \sqrt{\frac{-(m_\ell^{\tau\tau})^2-(m_\ell^{\mu\tau})^2+m_\tau^2}{m_\tau^2-m_\mu^2}}.
\label{Eq:calpha_salpha}
\end{eqnarray}
The diagonal masses are computed to be:

\begin{eqnarray}
m_e^2&=&(m_\ell^{ee})^2,
\nonumber \\
m_\mu^2&=&\frac{1}{2}
    \left[ (m_\ell^{\tau\tau})^2+(m_\ell^{\mu\mu})^2+(m_\ell^{\tau\mu})^2+(m_\ell^{\mu\tau})^2 - M^2\right],
\nonumber \\
m_\tau^2&=&\frac{1}{2}
    \left[ (m_\ell^{\tau\tau})^2+(m_\ell^{\mu\mu})^2+(m_\ell^{\tau\mu})^2+(m_\ell^{\mu\tau})^2 + M^2\right],
\label{Eq:mell_diagonalmass}
\end{eqnarray}
where
\begin{eqnarray}
M^4 &=& \left[ (m_\ell^{\tau\tau})^2-(m_\ell^{\mu\mu} )^2\right]^2
      + \left[ (m_\ell^{\tau\mu} )^2-(m_\ell^{\mu\tau})^2\right]^2
\nonumber \\ 
	&&+ 2(m_\ell^{\tau\tau}m_\ell^{\mu\tau}+m_\ell^{\mu\mu}m_\ell^{\tau\mu})^2
	  + 2(m_\ell^{\tau\tau}m_\ell^{\tau\mu}+m_\ell^{\mu\mu}m_\ell^{\mu\tau})^2.
\label{Eq:M4}
\end{eqnarray}
There are the following relations for non-diagonal and diagonal charged lepton masses:
\begin{eqnarray}
m_\ell^{\mu\mu}&=&S^2m_\tau + C^2m_\mu, \quad m_\ell^{\tau\tau}=C^2m_\tau + S^2m_\mu,
\nonumber \\
m_\ell^{\mu\tau}&=&\frac{1}{c_\beta^2-s_\alpha^2}
    \left[ (c_\alpha s_\alpha C^2 - c_\beta  s_\beta  S^2)m_\tau \right.
   -\left. (c_\beta  s_\beta  C^2 - c_\alpha s_\alpha S^2)m_\mu  \right],
\nonumber \\
m_\ell^{\tau\mu}&=&\frac{1}{c_\beta^2-s_\alpha^2}
    \left[ (c_\beta  s_\beta  C^2 - c_\alpha s_\alpha S^2)m_\tau \right.
   -\left. (c_\alpha s_\alpha C^2 - c_\beta  s_\beta  S^2)m_\mu  \right],
\label{Eq:mell_diag_to_nondiag}
\end{eqnarray}
where $C^2$ and $S^2$ are defined by 

\begin{eqnarray}
C^2=\frac{c_\alpha^2+c_\beta^2}{2}, \quad
S^2=\frac{s_\alpha^2+s_\beta^2}{2}.
\label{Eq:C2S2}
\end{eqnarray}
We use the hierarchical conditions of $\vert s_\alpha \vert, \vert s_\beta \vert \ll 1$ to realize the hierarchical mass pattern of $m_\mu \ll m_\tau$, namely, $m_\ell^{\mu\mu} \ll m_\ell^{\tau\tau}$.  This hierarchy in turn requires that

\begin{eqnarray}
&& \vert f_\mu^+\vert \ / \vert f_\tau^+\vert \sim \vert g_\mu^- \vert / \vert g_\tau^-\vert \sim m_\mu/m_\tau, \qquad
f_\mu^+ v_\rho \sim -g_\mu^- v_{\rho^\prime}, \qquad
f_\tau^+ v_\rho \sim g_\tau^- v_{\rho^\prime}.
\label{Eq:HierarchicalCoupling}
\end{eqnarray}
We consider that the fine-tuning of the charged lepton masses of Eq.(\ref{Eq:HierarchicalCoupling}) is the same level of the fine-tuning in the standard model.  To explain their hierarchical structure needs some other mechanisms, which we do not consider in this paper.  We only consider the permutation symmetry as a new symmetry behind neutrino oscillations once the charged lepton masses are consistently reproduced. 

It should be noted that our model induces dangerous flavor-changing interactions such as $\tau \rightarrow \mu\gamma, \mu\mu\mu, \mu ee$ mediated by $\eta$. In addition, the existence of $\rho$ and $\rho^\prime$ also induces these flavor-changing interactions because the charged leptons can simultaneously couple to two Higgs scalars, $\rho$ and $\rho^\prime$ \cite{FCNC}. Since the approximate $L_e$ conservation is satisfied by our interactions, all $L_e$-changing flavor interactions such as $\mu \rightarrow e \gamma$ including those mediated by $\eta$ can be well-suppressed. The $\rho$- and $\rho^\prime$-interactions are also found to be suppressed down to the phenomenologically acceptable level. The branching ratios of these processes are taken from Ref.\cite{PDG}:

\begin{itemize}
\item $B(\tau \rightarrow \mu \gamma) = \Gamma (\tau \rightarrow \mu \gamma)/\Gamma(\tau \rightarrow all) < 1.1 \times 10^{-6}$,

\item $B(\tau \rightarrow \mu \mu \mu) = \Gamma (\tau \rightarrow \mu\mu\mu)/\Gamma(\tau \rightarrow all) < 1.9 \times 10^{-6}$,

\item$B(\tau \rightarrow \mu ee) = \Gamma (\tau \rightarrow \mu ee)/\Gamma(\tau \rightarrow all) < 1.7 \times 10^{-6}$,
\end{itemize}
where $\Gamma(\tau \rightarrow all)$ ($< 2.3 \times 10^{-12}$ [GeV]) denotes the total decay width and $\Gamma (\tau \rightarrow \mu \gamma)$, $\Gamma (\tau \rightarrow  \mu\mu\mu)$ and $\Gamma (\tau \rightarrow \mu ee)$ denote the decay widths of $\tau \rightarrow \mu\gamma$, $\tau \rightarrow \mu\mu\mu$ and $\tau \rightarrow \mu e e$ processes, respectively. These decay widths are calculated to be, in the $\alpha \rightarrow 0$ limit,

\begin{eqnarray}
\Gamma (\tau \rightarrow \mu \gamma)&=&\frac{\alpha_{em} m_\tau^5}{96^2 \pi^4 \overline{m}_\rho^4}
     \frac{ (f_\mu^+)^4 + (g_\mu^-)^4 + (f_\tau^+ f_\mu^+)^2 + (g_\tau^- g_\mu^-)^2}{4},
\nonumber \\
\Gamma (\tau \rightarrow \mu\mu\mu)&=&\frac{1}{4}\frac{1}{8}\frac{m_\tau^5}{192 \pi^3 \overline{m}_\rho^4}
    \frac{(f_\mu^+)^4 + (g_\mu^-)^4}{4},
\nonumber \\
\Gamma (\tau \rightarrow \mu ee)&=&\frac{1}{4}\frac{1}{8}\frac{m_\tau^5}{192 \pi^3 m_\rho^4}
    \frac{(f_\mu^+ f_e)^2}{4},
\label{Eq:FCNCG}
\end{eqnarray}
where $\alpha_{em}$ is the fine-tuning constant and $\overline{m}_\rho$ is the averaged mass for $\rho$ and $\rho^\prime$. By using Eq.(\ref{Eq:MellElements}) and Eq.(\ref{Eq:FCNCG}) to relate the couplings to the charged lepton masses, we obtain the following constraints on the processes of $\tau \rightarrow \mu \gamma$, $\tau \rightarrow \mu \mu \mu$ and $\tau \rightarrow \mu ee$, respectively:

\begin{eqnarray}
\frac{m_\mu^2 (m_\tau^2+m_\mu^2)}{8 \overline{m}_\rho^4 v_\rho^4} < 1.74 \times 10^{-11} ~[{\rm GeV}^{-4}],
\\ \nonumber
\frac{m_\mu^4}{8 \overline{m}_\rho^4 v_\rho^4} < 4.70 \times 10^{-14} ~[{\rm GeV}^{-4}],
\\ \nonumber
\frac{m_\mu^2 m_e^2}{4 m_\rho^4 v_\rho^4} < 4.20 \times 10^{-14} ~[{\rm GeV}^{-4}],
\label{Eq:FCNCconst}
\end{eqnarray}
where we have used $v_\rho = v_\rho^\prime$ for simplicity. We, then, find that

\begin{enumerate}
\item for the $\tau \rightarrow \mu\gamma$ process mediated by $\rho,\rho^\prime$, $\overline{m}_\rho v_\rho>2.04 \times 10^2$ [GeV$^2$] from $B(\tau \rightarrow \mu \gamma) < 1.1 \times 10^{-6}$,

\item for the $\tau \rightarrow \mu\mu\mu$ process mediated by $\rho,\rho^\prime$, $\overline{m}_\rho v_\rho>1.28 \times 10^2$ [GeV$^2$] from $B(\tau \rightarrow \mu\mu\mu) < 1.9 \times 10^{-6}$,

\item for the $\tau \rightarrow \mu ee$ process mediated by $\rho$, $m_\rho v_\rho>7.69$ [GeV$^2$] from $B(\tau \rightarrow \mu ee) < 1.7 \times 10^{-6}$.
\end{enumerate}
Since we are anticipating that $m_{\rho,\rho^\prime} \sim v_{\rho,\rho^\prime} \sim v_{weak}$ is natural because $\rho$ and $\rho^\prime$ are related to the weak boson masses, thus these flavor-changing interactions are sufficiently suppressed.

The other interactions mediated by $\eta^\prime$ cause extra contributions on $\tau^- \rightarrow \mu^- \overline{\nu}_\mu \nu_\tau$, which is well described by the weak-boson-exchanges, and other flavor-changing modes induced by effects of $\sin \alpha \ne 0$ as well as $\sin \beta \ne 0$. These contributions may not be suppressed in practice; however, they become tiny because the coupling of $\eta^\prime$ to leptons is taken to be the same order of that of $\eta$ in the later discussions.

\section{\label{sec:4}Neutrino masses and oscillations}
Now, we are at the stage to estimating the neutrino masses and mixings in our model. The neutrino mass matrix is given by

\begin{eqnarray}
M_\nu= M_\nu^{tree}+M_\nu^{rad}
\label{Eq:Mnu331}
\end{eqnarray}
where $M_\nu^{tree}$ for the tree level mass matrix and $M_\nu^{rad}$ for the radiative mass matrix are parameterized by

\begin{eqnarray}
M_\nu^{tree}=\left(
    \begin{array}{ccc}
    0 & 0                         &  0 \\
    0 & 1 & -1 \\
    0 &  -1 &  1          \\
    \end{array}
    \right) m_\nu^{tree}, \quad
M_\nu^{rad}=
   \left( 
    \begin{array}{ccc}
    \delta m_\nu^{ee}                              & 
	 \delta m_\nu^{e\mu }&
	 \delta m_\nu^{e\tau} \\
     \delta m_\nu^{e\mu}&
	\delta m_\nu^{\mu\mu}                          & 
	 \delta m_\nu^{\mu\tau} \\
     \delta m_\nu^{e\tau}&
	 \delta m_\nu^{\mu\tau}&
	\delta m_\nu^{\tau\tau} \\
    \end{array}
    \right).
\label{Eq:MnuMassMatrix0}
\end{eqnarray}
The tree-level mass is calculated to be $m_\nu^{tree} = g_s^+ v_s$, which is supplied by the interaction of $g_s^+ \overline{\left( \psi_L^- \right)^c} s \psi_L^-$ via the type II seesaw mechanism \cite{type2seesaw} using the $\eta s \eta$ and $\eta^\prime s \eta^\prime$ as already mentioned. The smallness of the tree level neutrino masses is explained by the smallness of the $v_s$, which is estimated to be $v_s \sim \left( \mu_1 v_\eta^2 + \mu_2 v_{\eta^\prime}^2 \right)/(2m_s^2)$, leading to $v_s \ll v_{\eta,\eta^\prime} \sim v_{weak}$ for $\mu_{1,2} \sim m_s \gg v_{\eta,\eta^\prime}$. The radiatively generated masses of $\delta m_\nu$'s are calculated below.  To do so, we further divide $M_\nu^{rad}$ into two parts: 
\begin{eqnarray}
M_\nu^{rad}=M_\nu^C+M_\nu^H,
\label{Eq:MnuRadiative}
\end{eqnarray}
where $M_\nu^C$ and $M_\nu^H$ correspond to the charged lepton mediated one-loop diagrams (FIG.\ref{Fig:ChargedOneLoops}) and the heavy lepton mediated one-loop diagrams (FIG.\ref{Fig:HeavyOneLoops}), respectively. There are other possible loop diagrams of FIG.\ref{Fig:SuppressedLoops}; however, the contributions from these diagrams are well suppressed by the large mass ($\gg v_{weak}$) arising from the propagator of $s$. Since our charged lepton mass matrix is transformed into $M_\ell^{diag} =$ $diag.(m_e,m_\mu,m_\tau) = U_\ell^\dagger M_\ell V_\ell$, the neutrino mass matrix, $M_\nu$, is also transformed into a matrix spanned by the weak base, $M_\nu^{weak}$, defined by

\begin{eqnarray}
M_\nu^{weak} \equiv U_\ell^\dagger M_\nu U_\ell = (M_\nu^{tree})^{weak} + (M_\nu^C)^{weak}+(M_\nu^H)^{weak}
\label{Eq:MnuWeak}
\end{eqnarray}
with $(M_\nu^a)^{weak}=U_\ell^\dagger M_\nu^a U_\ell$, where $a$ stands for $tree$, $C$ and $H$.

The mass matrix $(M_\nu^{tree})^{weak}$ is simply given by:

\begin{eqnarray}
(M_\nu^{tree})^{weak}=\left(
    \begin{array}{ccc}
    0 & 0                         &  0 \\
    0 & (c_\alpha + s_\alpha)^2   & -c_\alpha^2 + s_\alpha^2 \\
    0 &  -c_\alpha^2 + s_\alpha^2 & (c_\alpha - s_\alpha)^2           \\
    \end{array}
    \right) m_\nu^{tree}.
\label{Eq:UMnuTreeU}
\end{eqnarray}
At a first glance, this deviation due to $\sin\alpha\neq 0$ may yield $U_{e3}$ through $\epsilon \neq 0$ as in Eq.(\ref{Eq:Ue3}), which is given by $\epsilon$=$(c_\alpha - s_\alpha)^2-[(c_\alpha + s_\alpha)^2+(\sigma^{-1}-\sigma)(-c_\alpha^2 + s_\alpha^2)]$. However, one observes that $\epsilon$=0 for any values of $\sin\alpha$ by the use of $\sigma$ to be obtained in Eq.(\ref{Eq:sigmaFromHeavy}) from heavy lepton contributions, which will be found to much dominate over the corresponding charged lepton contributions. We safely state that the tree level contributions to $U_{e3}$ are negligible. 

The mass matrix $(M_\nu^C)^{weak}$ induced by the charged lepton exchanges is expressed as

\begin{eqnarray}
(M_\nu^C)^{weak}=
   \left( 
    \begin{array}{ccc}
    \delta m_\nu^{Cee}                              & 
	 \delta m_\nu^{Ce\mu }&
	 \delta m_\nu^{Ce\tau} \\
     \delta m_\nu^{Ce\mu}&
	\delta m_\nu^{C\mu\mu}                          & 
	 \delta m_\nu^{C\mu\tau} \\
     \delta m_\nu^{Ce\tau}&
	 \delta m_\nu^{C\mu\tau}&
	\delta m_\nu^{C\tau\tau} \\
    \end{array}
    \right)
=
   \left( 
    \begin{array}{ccc}
    2\delta C_\nu^{ee}                              & 
	 \delta C_\nu^{e\mu }   + \delta C_\nu^{\mu  e} &
	 \delta C_\nu^{e\tau}   + \delta C_\nu^{\tau e}   \\
     \delta C_\nu^{e\mu}    + \delta C_\nu^{\mu  e} &
	2\delta C_\nu^{\mu\mu}                          & 
	 \delta C_\nu^{\mu\tau} + \delta C_\nu^{\tau\mu} \\
     \delta C_\nu^{e\tau}   + \delta C_\nu^{\tau e} &
	 \delta C_\nu^{\mu\tau} + \delta C_\nu^{\tau\mu} &
	2\delta C_\nu^{\tau\tau} \\
    \end{array}
    \right),
\label{Eq:MnuChargedWeak}
\end{eqnarray}
where $\delta C_\nu^{ij}$ ($i,j=e,\mu,\tau$) denotes the radiatively induced neutrino masses corresponding to FIG.\ref{Fig:ChargedOneLoops}:

\begin{eqnarray}
\delta C_\nu^{ij}=\left(
    U_\ell^\dagger {\bf f}_\ell U_\ell M_\ell^{diag} V_\ell^\dagger {\bf g}_\ell {\bf F}_\ell U_\ell
    \right)^{ij},
\label{Eq:Cmuij}
\end{eqnarray}
which can be calculated as 

\begin{eqnarray}
{\bf f}_\ell=\left(
    \begin{array}{ccc}
    0                      &  \lambda_4 f_{e\mu}/2   & \lambda_4 f_{e\tau}/2 \\
    -\lambda_4 f_{e\mu}/2  &  0                      & \lambda_3 f_{\mu\tau} \\
    -\lambda_4 f_{e\tau}/2 &  -\lambda_3 f_{\mu\tau} & 0           \\
    \end{array}
    \right),
\quad	
{\bf g}_\ell=\frac{1}{\sqrt{2}} 
    \left(
    \begin{array}{ccc}
    \sqrt{2}f_e &  0       & 0\\
    0           & -g_\mu^- & -g_\tau^- \\
    0           &  g_\mu^- &  g_\tau^-       \\
    \end{array}
    \right),
\label{Eq:fellgell}
\end{eqnarray}

\begin{eqnarray}
{\bf F}_\ell = v_\eta v_\chi \left(
    \begin{array}{ccc}
    F(m_e^2, m_\eta^2, m_\rho^2) & 0                              & 0 \\
    0                            & F(m_\mu^2, m_\eta^2, m_\rho^2) & 0 \\
    0                            & 0                              & F(m_\tau^2, m_\eta^2, m_\rho^2)\\
	\end{array}
	\right)
	\equiv v_\eta v_\chi\left(
    \begin{array}{ccc}
    F_\ell^e & 0          & 0 \\
    0        & F_\ell^\mu & 0 \\
    0        & 0          & F_\ell^\tau \\
	\end{array}
	\right)
\label{Eq:Fell}
\end{eqnarray}
with

\begin{eqnarray}
F(x,y,z)&=&\frac{1}{16\pi^2}
\left[
\frac{x\ln x}{(x-y)(x-z)}+\frac{y\ln y}{(y-x)(y-z)}+\frac{z\ln z}{(z-y)(z-x)}
\right].
\label{Eq:F}
\end{eqnarray}
After some calculations, we obtain:

\begin{eqnarray}
\delta C_\nu^{ee}&=&0 ,
\nonumber \\
\delta C_\nu^{e\mu} &=& \frac{1}{2}\lambda_4 v_\eta v_\chi
    \left[
        m_\mu(c_\alpha f_{e\mu} - s_\alpha f_{e\tau})
        (-g_\mu^- c_\alpha c_\beta F_\ell^\mu + g_\tau^- s_\alpha c_\beta F_\ell^\tau
		 -g_\mu^- c_\alpha s_\beta  F_\ell^\mu + g_\tau^- s_\alpha s_\beta F_\ell^\tau)
	\right.
\nonumber \\
    &&\left.
      + m_\tau(s_\alpha f_{e\mu} + c_\alpha f_{e\tau})
        (-g_\mu^- c_\alpha s_\beta F_\ell^\mu + g_\tau^- s_\alpha s_\beta F_\ell^\tau
		 +g_\mu^- c_\alpha c_\beta F_\ell^\mu - g_\tau^- s_\alpha c_\beta F_\ell^\tau)
    \right],
\nonumber \\
\delta C_\nu^{e\tau}&=&\frac{1}{2}\lambda_4 v_\eta v_\chi
    \left[
        m_\mu(c_\alpha f_{e\mu} - s_\alpha f_{e\tau})
        (-g_\mu^- s_\alpha c_\beta F_\ell^\mu - g_\tau^- c_\alpha c_\beta F_\ell^\tau
		 -g_\mu^- s_\alpha s_\beta F_\ell^\mu - g_\tau^- c_\alpha s_\beta F_\ell^\tau)
    \right.
\nonumber \\
    &&\left.
      + m_\tau(s_\alpha f_{e\mu} + c_\alpha f_{e\tau})
        (-g_\mu^- s_\alpha s_\beta F_\ell^\mu - g_\tau^- c_\alpha s_\beta F_\ell^\tau
		 -g_\mu^- s_\alpha c_\beta F_\ell^\mu - g_\tau^- c_\alpha c_\beta F_\ell^\tau)
    \right],
\nonumber \\
\delta C_\nu^{\mu e}&=& \frac{1}{\sqrt{2}}\lambda_4 f_e v_\eta v_\chi m_e
    (-c_\alpha f_{e\mu} + s_\alpha f_{e\tau}) F_\ell^e,
\nonumber \\
\delta C_\nu^{\mu \mu}&=&\frac{1}{\sqrt{2}}\lambda_3 f_{\mu\tau} v_\eta v_\chi m_\tau 
        (-g_\mu^- c_\alpha s_\beta F_\ell^\mu + g_\tau^- s_\alpha s_\beta F_\ell^\tau
		 +g_\mu^- c_\alpha c_\beta F_\ell^\mu - g_\tau^- s_\alpha c_\beta F_\ell^\tau),
\nonumber \\
\delta C_\nu^{\mu \tau}&=&\frac{1}{\sqrt{2}}\lambda_3 f_{\mu\tau} v_\eta v_\chi m_\tau   
        (-g_\mu^- s_\alpha s_\beta F_\ell^\mu - g_\tau^- c_\alpha s_\beta F_\ell^\tau
		 -g_\mu^- s_\alpha c_\beta F_\ell^\mu - g_\tau^- c_\alpha c_\beta F_\ell^\tau),
\nonumber \\
\delta C_\nu^{\tau e}&=& \frac{1}{\sqrt{2}}\lambda_4 f_e v_\eta v_\chi m_e
    (-s_\alpha f_{e\mu} - c_\alpha f_{e\tau}) F_\ell^e,
\nonumber \\
\delta C_\nu^{\tau \mu}&=&-\frac{1}{\sqrt{2}}\lambda_3 f_{\mu\tau} m_\mu 
    (-g_\mu^- c_\alpha c_\beta F_\ell^\mu + g_\tau^- s_\alpha c_\beta F_\ell^\tau
	 -g_\mu^- c_\alpha s_\beta  F_\ell^\mu + g_\tau^- s_\alpha s_\beta F_\ell^\tau),
\nonumber \\
\delta C_\nu^{\tau \tau}&=&-\frac{1}{\sqrt{2}}\lambda_3 f_{\mu\tau} v_\eta v_\chi m_\mu 
    (-g_\mu^- s_\alpha c_\beta F_\ell^\mu - g_\tau^- c_\alpha c_\beta F_\ell^\tau
	 -g_\mu^- s_\alpha s_\beta F_\ell^\mu - g_\tau^- c_\alpha s_\beta F_\ell^\tau).
\nonumber \\
\label{Eq:Cij}
\end{eqnarray}

For the heavy lepton contributions, we take into account the rotation from the mass eigenstates $\vert \psi^e, \psi^-, \psi^+ \rangle$ (appearing in the calculations) to the flavor eigenstates $\vert \psi^e, \psi^\mu, \psi^\tau \rangle$ (appearing in the neutrino mass matrix) by introducing $S$ as $\vert \psi^e, \psi^-, \psi^+ \rangle$ = $S\vert \psi^e, \psi^\mu, \psi^\tau \rangle$. The mass matrix of $(M_\nu^H)^{weak}$ is, then, computed to be:

\begin{eqnarray}
(M_\nu^H)^{weak}=
   \left( 
    \begin{array}{ccc}
    \delta m_\nu^{Hee}                              & 
	 \delta m_\nu^{He\mu }&
	 \delta m_\nu^{He\tau} \\
     \delta m_\nu^{He\mu}&
	\delta m_\nu^{H\mu\mu}                          & 
	 \delta m_\nu^{H\mu\tau} \\
     \delta m_\nu^{He\tau}&
	 \delta m_\nu^{H\mu\tau}&
	\delta m_\nu^{H\tau\tau} \\
    \end{array}
    \right)
=
    \left( 
    \begin{array}{ccc}
    2\delta H_\nu^{ee}                              & 
	 \delta H_\nu^{e\mu }   + \delta H_\nu^{\mu  e} &
	 \delta H_\nu^{e\tau}   + \delta H_\nu^{\tau e}   \\
     \delta H_\nu^{e\mu}    + \delta H_\nu^{\mu  e} &
	2\delta H_\nu^{\mu\mu}                          & 
	 \delta H_\nu^{\mu\tau} + \delta H_\nu^{\tau\mu} \\
     \delta H_\nu^{e\tau}   + \delta H_\nu^{\tau e} &
	 \delta H_\nu^{\mu\tau} + \delta H_\nu^{\tau\mu} &
	2\delta H_\nu^{\tau\tau} \\
    \end{array}
    \right),
\label{Eq:MnuChargedWeak}
\end{eqnarray}
where $\delta H_\nu^{ij}$ ($i,j=e,\mu,\tau$) denotes the radiatively induced neutrino masses corresponding to FIG.\ref{Fig:HeavyOneLoops}, which can be calculated as

\begin{eqnarray}
\delta H_\nu^{ij}=\left(
    U_\ell^\dagger S^\dagger {\bf f}_\ell M_\kappa^{diag} {\bf f}_\kappa {\bf F}_\kappa S U_\ell
    \right)^{ij},
\label{Eq:Hmuij}
\end{eqnarray}
where

\begin{eqnarray}
{\bf f}_\ell=\left(
    \begin{array}{ccc}
    0                      &  0                     &  \lambda_3 f_{e\ell}/2 \\
    0                      &  0                     & -\lambda_4 f_{\mu\tau} \\
    -\lambda_3 f_{e\ell}/2 &  \lambda_4 f_{\mu\tau} & 0           \\
    \end{array}
    \right),
\quad	
{\bf f}_\kappa=\left(
    \begin{array}{ccc}
    f_\kappa^e   &  0           & 0          \\
    0            &  g_\kappa^+  & 0          \\
    0            &  0           & f_\kappa^+ \\
    \end{array}
    \right),
\label{Eq:f_kappa}
\end{eqnarray}

\begin{eqnarray}
{\bf F}_\kappa =  v_\eta v_\rho \left(
    \begin{array}{ccc}
    F((m_\kappa^e)^2, m_\eta^2, m_\chi^2) & 0                   & 0 \\
    0                              & F((m_\kappa^-)^2, m_\eta^2, m_\chi^2) & 0 \\
    0                              & 0                          & F((m_\kappa^+)^2, m_\eta^2, m_\chi^2)\\
	\end{array}
	\right)
	\equiv  v_\eta v_\rho \left(
    \begin{array}{ccc}
    F_\kappa^e & 0          & 0 \\
    0          & F_\kappa^- & 0 \\
    0          & 0          & F_\kappa^+ \\
	\end{array}
	\right),
\label{Eq:Fkappa}
\end{eqnarray}

\begin{eqnarray}
M_\kappa^{diag}=\left(
    \begin{array}{ccc}
    m_\kappa^e   &  0           & 0          \\
    0            &  m_\kappa^-  & 0          \\
    0            &  0           & m_\kappa^+ \\
    \end{array}
    \right),
\quad
S=\frac{1}{\sqrt{2}}\left(
    \begin{array}{ccc}
    \sqrt{2} &  0 & 0 \\
    0        & -1 & 1 \\
    0        &  1 & 1 \\
    \end{array}
    \right).
\label{Eq:M_kappa_S_kappa}
\end{eqnarray}
After some calculations, we obtain:

\begin{eqnarray}
\delta H_\nu^{ee}&=&0,
\nonumber \\ 
\delta H_\nu^{e\mu}&=&\frac{1}{\sqrt{2}}(c_\alpha - s_\alpha ) \lambda_3 f_{e\ell}v_\eta v_\rho f_\kappa^+ m_\kappa^+ F_\kappa^+,
\nonumber \\ 
\delta H_\nu^{e\tau}&=&\frac{1}{\sqrt{2}}(c_\alpha + s_\alpha ) \lambda_3 f_{e\ell}v_\eta v_\rho f_\kappa^+ m_\kappa^+F_\kappa^+,
\nonumber \\ 
\delta H_\nu^{\mu e}&=&-\frac{1}{\sqrt{2}}(c_\alpha - s_\alpha ) \lambda_3 f_{e\ell}v_\eta v_\rho f_\kappa^e m_\kappa^e F_\kappa^e,
\nonumber \\ 
\delta H_\nu^{\mu \mu}&=&\frac{1}{2}(c_\alpha^2 - s_\alpha^2 ) \lambda_4 f_{\mu\tau}v_\eta v_\rho 
    (f_\kappa^+ m_\kappa^+ F_\kappa^+ - g_\kappa^+ m_\kappa^- F_\kappa^-),
\nonumber \\ 
\delta H_\nu^{\mu \tau}&=&\frac{1}{2} \lambda_4 f_{\mu\tau}v_\eta v_\rho 
    \left[(c_\alpha + s_\alpha )^2 f_\kappa^+ m_\kappa^+ F_\kappa^+ 
	    + (c_\alpha - s_\alpha )^2 g_\kappa^+ m_\kappa^- F_\kappa^-
    \right],
\nonumber \\
\delta H_\nu^{\tau e}&=&-\frac{1}{\sqrt{2}}(c_\alpha + s_\alpha ) \lambda_3 f_{e\ell}v_\eta v_\rho f_\kappa^e m_\kappa^e F_\kappa^e,
\nonumber \\ 
\delta H_\nu^{\tau \mu}&=&\frac{1}{2}\lambda_4 f_{\mu\tau}v_\eta v_\rho 
    \left[-(c_\alpha - s_\alpha )^2 f_\kappa^+ m_\kappa^+ F_\kappa^+ 
	    - (c_\alpha + s_\alpha )^2 g_\kappa^+ m_\kappa^- F_\kappa^-
    \right],
\nonumber \\
\delta H_\nu^{\tau \tau}&=&-\delta H_\nu^{\mu \mu}.
\label{Eq:Hij}
\end{eqnarray}

There are two contributions from $(M_\nu^C)^{weak}$ and $(M_\nu^H)^{weak}$ to radiatively induced neutrino masses. For $\nu_e$-$\nu_\mu$ and $\nu_e$-$\nu_\tau$ contributions, the contributions from $(M_\nu^C)^{weak}$ are found to be always much smaller than those from $(M_\nu^H)^{weak}$. In this case, there is a relation of 

\begin{eqnarray}
\sigma=-\frac{1+\tan \alpha}{1-\tan \alpha}
\label{Eq:sigmaFromHeavy}
\end{eqnarray}
calculated from Eq.(\ref{Eq:Hij}), leading to

\begin{eqnarray}
\sin^2 2\theta_{atm}=4\frac{\sigma^2}{(1+\sigma^2)^2}.
\label{Eq:sin2atmFromHeavy}
\end{eqnarray}

To see how neutrino masses and mixings arise, we discuss the simplest case with $\sin \alpha=0$, where the charged-lepton contributions are neglected. The full estimation including the case of $\sin \alpha \neq 0$ is to be preformed by numerical calculations, where the charged-lepton contributions are properly taken into account. The neutrino mass matrix turns out to be:

\begin{eqnarray}
M_\nu^{weak} = \left(
    \begin{array}{ccc}
    0             &  H_\nu^{e\mu}                 & H_\nu^{e\tau} \\
    H_\nu^{e\mu}  &  m_\nu^{tree}+H_\nu^{\mu\mu}  & -m_\nu^{tree} \\
    H_\nu^{e\tau} & -m_\nu^{tree}                 & m_\nu^{tree} + H_\nu^{\tau\tau}\\
    \end{array}
    \right),
\label{Eq:Mnu_alphaZero}
\end{eqnarray}
where $H_\nu^{\mu\tau}$ is absent because of $H_\nu^{\mu\tau} = \delta H_\nu^{\mu\tau} + \delta H_\nu^{\tau\mu} = 0$ for $\sin \alpha=0$, and 

\begin{eqnarray}
H_\nu^{e\mu}&=&H_\nu^{e\tau}=\frac{\lambda_3 f_{e\ell} v_\eta v_\rho}{2 v_\chi}
    \left[ 
	(m_\kappa^+)^2 F((m_\kappa^+)^2,m_\eta^2,m_\chi^2) - (m_\kappa^e)^2 F((m_\kappa^e)^2,m_\eta^2,m_\chi^2) 
	\right],
\nonumber \\
H_\nu^{\mu\mu}&=&-H_\nu^{\tau\tau}=\frac{\lambda_4 f_{\mu\tau} v_\eta v_\rho}{v_\chi}
    \left[ 
	(m_\kappa^+)^2 F((m_\kappa^+)^2,m_\eta^2,m_\chi^2) - (m_\kappa^-)^2 F((m_\kappa^-)^2,m_\eta^2,m_\chi^2) 
	\right],
\label{Eq:simpleDeltaMnu}
\end{eqnarray}
where $m_\kappa^+$=$f_\kappa^+ \chi$ and $m_\kappa^e$=$f_\kappa^e\chi$ are used to eliminate $f_\kappa$'s. These structures are simply explained by the form of effective operators.  The equation of $H_\nu^{e\mu}=H_\nu^{e\tau}$ is due to the appearance of $\psi_L^e\psi_L^+$ giving rise to $\nu_e\nu_\mu+\nu_e\nu_\tau$ while that of $H_\nu^{\mu\mu}=-H_\nu^{\tau\tau}$ is due to the appearance of $\psi_L^-\psi_L^+$ giving rise to $\nu_\tau\nu_\tau-\nu_\mu\nu_\mu$. This direct correspondence of the signs of the mass terms is partially owing to the $S_{2L}$-conserved heavy lepton interactions with $\chi$ even after the spontaneous breaking. Then, our mass matrix of Eq.(\ref{Eq:Mnu}) has the following mass parameters:

\begin{eqnarray}
a&=&0, \quad b=H_\nu^{e\mu}, \quad c=H_\nu^{e\mu}, \quad d=m_\nu^{tree}+H_\nu^{\mu\mu},
\nonumber \\
e&=&-m_\nu^{tree}, \quad f=m_\nu^{tree} - H_\nu^{\mu\mu}
\label{Eq:massparameter}.
\end{eqnarray}

From these mass parameters, we find that our model naturally derives the maximal atmospheric neutrino mixing because $\sigma = -c/b = -1$. We mention that, in the original Zee model, the fine-tuning of lepton number violating couplings, which is characterized by ``inverse hierarchy'' in the couplings such as $f_{e\tau}m_\tau^2 \sim f_{e\mu}m_\mu^2$ \cite{inverseHierarchy} is necessary to yield bimaximal mixing structure. However, in our $SU(3)_L \times U(1)_N$ case, the bimaximal structure is caused by the dominance of the heavy lepton contributions. Also, we find that our model is capable of explaining the large solar neutrino mixing with $\sin^2 2\theta_\odot \sim 0.8$ because $H_\nu^{\mu\mu}=\mathcal{O}(H_\mu^{e\mu})$ is a reasonable expectation for $f_{e\mu} \sim f_{\mu\tau}$, $\lambda_3 \sim \lambda_4$ and $m_\kappa^e \sim m_\kappa^-$. We then observe that, in the case of $\sin \alpha=0$,

\begin{enumerate}
\item The appearance of the maximal atmospheric neutrino mixing is ensured by the contributions characterized by $H_\nu^{e\mu}=H_\nu^{e\tau}$. It is mainly because the heavy lepton interactions with $\chi$ respects $S_{2L}$ even after the spontaneous breaking, which ensures $H_\nu^{e\mu}=H_\nu^{e\tau}$ giving $\delta m_\nu^{He\mu}$ = $\delta m_\nu^{He\tau}$ by $\mu \leftrightarrow \tau$ as noted. 
\item The deviation from the maximal solar neutrino mixing is due to the contributions characterized by $H_\nu^{\mu\mu} \sim H_\nu^{e\mu}$, which roughly suggests that $m_\kappa^- \sim m_\kappa^e$.
\end{enumerate}
The (suppressed) charged lepton contributions of Eq.(\ref{Eq:Cij}) give, for $\sin\alpha=0$ together with $\sin\beta =  0$, the relevant radiative masses of $\delta m_\nu^{Ce\mu,e\tau}$ to be:
\begin{eqnarray}\label{ChargedAtm}
\delta m_\nu^{Ce\mu} (=\delta C^{e\mu}+\delta C^{\mu e}) &\approx& \frac{1}{2}\lambda_4 f_{e\tau}g_\mu^- v_\eta v_\chi m_\tau  F_\ell^\mu, 
\nonumber \\
\delta m_\nu^{Ce\tau} (=\delta C^{e\tau}+\delta C^{\tau e}) &\approx& -\frac{1}{2}\lambda_4 f_{e\tau}g_\tau^- v_\eta v_\chi m_\tau  F_\ell^\tau.
\end{eqnarray}
Because $F_\ell^\mu\sim F_\ell^\tau$ for $m_\eta^2\sim m_\rho^2 \gg m_{e,\mu,\tau}^2$, we observe from Eq.(\ref{Eq:HierarchicalCoupling}) that $\vert \delta m_\nu^{Ce\mu} / \delta m_\nu^{Ce\tau}\vert\approx \vert g^-_\mu / g^-_\tau \vert \sim m_\mu/m_\tau$.  This relation disfavors the maximal atmospheric mixing.  In order to estimate the effects of $\sin \alpha \neq 0$, we next perform numerical estimation including these charged lepton contributions.
\section{\label{sec:5}Estimation of neutrino masses}
In this section, to see whether our model can really reproduce the (nearly) maximal atmospheric neutrino mixing of $\sin^2 2\theta_{atm} \sim 1.0$ and the LMA solution with $\sin^2 2\theta_\odot \sim 0.8$, we now perform the numerical calculations for $\sin \alpha=0.0, 0.1, 0.2$. We take the following assumptions on the magnitudes of various parameters: 

\begin{enumerate}
\item Since $v_{\eta,\eta^\prime}$ and $v_{\rho,\rho^\prime}$ are related to weak boson masses proportional to $v_{weak}^2 = \sum_{all} v_{Higgs}^2$, we put $v_\eta = v_{\eta^\prime} = v_{weak}/20$, $v_\rho = v_{\rho^\prime}= v_{weak}/\sqrt{2}$ where $v_{weak}=(2\sqrt{2}G_F)^{-1/2} = 174$ GeV.

\item $v_\chi$ is a source of masses for the heavy leptons, exotic quarks and exotic gauge bosons. Also $\chi$ is the key field for the symmetry breaking of $SU(3)_L \times U(1)_N \rightarrow SU(2)_L \times U(1)_Y$, leading to $v_\chi \gg v_{weak}$ and we put $v_\chi = 10v_{weak}$.

\item The masses of the triplet Higgs scalars $\eta,\eta^\prime,\rho,\rho^\prime$ are set to be $m_\eta=m_{\eta^\prime}=m_\rho=m_{\rho^\prime}=v_{weak}$ and the mass of $\chi$ is set to be $m_\chi=v_\chi$.

\item The mass of heavy lepton, $\kappa^e$, is taken to be $m_\kappa^e = 2$ TeV and the masses of $\kappa^\pm$ are varied within the range of $1.0$ TeV $\le m_\kappa^\pm \le 10.0$ TeV.

\item In order to realize the observed value of $\Delta m_{atm}^2=3.0\times10^{-3}$eV$^2$ by the type II seesaw mechanism, we take $m_s=4.6 \times 10^9 v_{weak}$ with $g_s^+ = e$ in Eq.(\ref{Eq:Yukawa}) and $\mu_1=\mu_2=m _s$ in Eq.(\ref{Eq:higgsV}), where $e$ stands for the electromagnetic coupling.

\item The couplings $f_{ij} (i,j=e,\mu,\tau)$ and $\lambda_{3,4}$ are taken to favor the large solar neutrino mixing angle, where $f_{e\ell} = 1.0 \times 10^{-8}$ and $f_{\mu\tau}=f_{e\ell}/2$ together with $\lambda_3=\lambda_4(=0.3)$ to focus $m_\kappa^- \sim m_\kappa^e$. The smallness of the Yukawa couplings $f_{e\ell}$ is taken to be "natural"  because the limit of $f_{e\ell} \rightarrow 0$ enhances the symmetry of the theory, $L_e$ symmetry, as mentioned above.
\end{enumerate}
It may be helpful to list these parameter values, which are summarized in TABLE \ref{Tab:FixedParameters}. We note again that 
\begin{itemize}
\item $v_{\eta,\eta^\prime,\rho,\rho^\prime}$ are so chosen to reproduce weak boson masses, 
\item $v_\chi$ is so chosen to give larger masses of at least 1 TeV to exotic particles,
\item the masses of $\eta,\eta^\prime,\rho,\rho^\prime,\chi$ are so chosen near these VEV's,
\item the tiny value of $f_{e\ell}$ enjoys the t'Hooft's "naturalness". 
\end{itemize}
These values are selected by the considerations within the $SU(3)_L \times U(1)_N$ framework. On the other hand, the remaining parameters are taken to accommodate the current neutrino observation data, e.g., 
\begin{itemize}
\item the mass of the sextet scalar $s$ and the related couplings $g_s^+, \mu_{1,2}$ are so chosen to reproduce $\Delta m_{atm}^2$ via the type II seesaw mechanism,
\item $f_{e\ell,\mu\tau}$ and $\lambda_{3,4}$ are so chosen to satisfy $\lambda_3f_{e\ell}/2$ = $\lambda_4f_{\mu\tau}$ (to enhance $m_\kappa^- \sim m_\kappa^e$) for the large solar neutrino mixing as in Eq.(\ref{Eq:simpleDeltaMnu}).
\end{itemize}
After taking these values, we have $m_\kappa^{e,\pm}$ as the free parameters. For the present analysis, we show various results with $m_\kappa^e$=2 TeV because the essence of our analysis is not altered for other similar values of $m_\kappa^e$. Now, our aim in this section is to give the numerical demonstration that our model with these parameters can really reproduce the observed neutrino oscillation data.  

For the sake of simplicity, we choose $\beta$ = $\alpha$ to estimate the charged lepton contributions of Eq.(\ref{Eq:Cij}).  We search the allowed region, where the following conditions 

\begin{enumerate}
\item $\Delta m_{atm}^2 = (3.0 \pm 0.01) \times 10^{-3}$ eV$^2$ for atmospheric neutrino oscillations,
\item $\sin^2 2 \theta_\odot = 0.6-0.8$ for solar neutrino oscillations,
\end{enumerate}
are satisfied.  The allowed region is selected by imposing the constraints of $\Delta m_\odot^2 \sim 10^{-5}-10^{-4}$ eV$^2$ and $\sin^2 2\theta_{atm}^{obs} \sim 0.9-1.0$ for $\theta_{atm}^{obs}=\theta_{atm}+\xi_{23}$ as in Eq.(\ref{Eq:xi12_xi23}). Note that, for the heavy lepton contributions alone, $\sin^2 2\theta_{atm} \geq 0.9$ is found to yield $\sin \alpha \leq 0.16$ from Eq.(\ref{Eq:sin2atmFromHeavy}).

The nearly maximal atmospheric neutrino mixing, $\sin^2 2\theta_{atm}^{obs} \mapgeq 0.9$, is naturally realized as shown in FIG.\ref{Fig:Sin} with $\sin \alpha \mapleq 0.1$.  This is because the heavy lepton contributions provides $\delta m_\nu^{e\mu}$ $\sim$ $\delta m_\nu^{e\tau}$ dominated by the $S_{2L}$-preserving contributions.  Although the charged lepton contributions generate $\delta m_\nu^{Ce\mu} / \delta m_\nu^{Ce\tau} \sim m_\mu/m_\tau$, which spoils the presence of the maximal atmospheric neutrino mixing, these contributions turn out to be much suppressed as shown in FIG.\ref{Fig:NuMassesC_H}. The radiative masses of $\delta m_\nu^{e\mu}$ and $\delta m_\nu^{e\tau}$ are essentially controlled by heavy lepton contributions, providing the (almost) maximal atmospheric neutrino mixing.

The large solar neutrino mixing calls for the similar magnitude of the radiatively induced neutrino masses. Namely, at least one of $\delta m_\nu^{\mu\mu,\mu\tau,\tau\tau}$ is the same order of magnitude as $\delta m_\nu^{e\mu,e\tau}$.  This condition is found to be realized by the appearance of $m_\kappa^-$ confined around 3$-$4 TeV as shown in FIG.\ref{Fig:HeavyMasses}, where the allowed region of $m_\kappa^\pm$ is examined. This behavior of $m_\kappa^-$ reflects the naive expectation of $m_\kappa^- \sim m_\kappa^e$ for $\sin \alpha=0$. In this region, the relevant radiatively induced neutrino masses are kept almost the same as shown in FIG.\ref{Fig:NuMasses} evaluated at $\sin\alpha = 0$ for demonstration. Thus, the large solar neutrino mixing is indeed possible to occur owing to the condition of Eq.(\ref{Eq:conditionOfsin08}) and $\sin^2 2\theta_\odot^{obs} = 0.67-0.98$ corresponding to 
\begin{equation}\label{SolarTan}
\tan^2\theta_\odot^{obs}=0.27-0.75
\end{equation}
are obtained because of the positive corrections by $\xi_{12}$ in $\theta_\odot^{obs}=\theta_\odot+\xi_{12}$. This region lies within the experimentally allowed region of $\tan^2\theta_\odot=0.24-0.89$ \cite{RecentSolar}. As shown in FIG.\ref{Fig:Xi}, the deviations of $\xi$'s remain suppressed to be: $\xi_{12(23)}/\theta_{12(23)}\sim 0.1$ for $\sin\alpha\mapleq 0.1$; however, in the region of $\sin\alpha\mapgeq 0.1$, these deviations become larger than ${\mathcal{O}}(0.1)$ of the original mixing angles, whose magnitudes may exceed the perturbative regime, where these deviations have been calculated. Hereafter, we do not specify the superscript of $obs$.  These deviations shift the mixing angles upwards. 

The charged lepton contributions present in the radiative masses of $\delta m_\nu^{\mu\mu,\mu\tau,\tau\tau}$ give favorable effects on $\sin^2 2\theta_\odot$ that reduce the magnitude of $\sin^2 2\theta_\odot$ from unity. So, we need not worry about the charged lepton contributions on these three masses. The effects of the charged lepton contributions in the $\mu$-$\tau$ sector arise in the magnitude of $U_{e3}$, which is shown in FIG.\ref{Fig:Ue3C_H}. Further shown in FIG.\ref{Fig:Ue3} is the sum of the charged lepton and heavy lepton contributions to $U_{e3}$, which turns out to be smaller than the experimental upper bound. 

Finally, in FIG.\ref{Fig:M1m2m3} and FIG.\ref{Fig:DeltaM}, we, respectively, show the neutrino mass eigenvalues and the squared mass differences $\Delta m_{atm}^2$ and $\Delta m_\odot^2$. These observations show that our model has the capability of explaining the observed properties of the atmospheric neutrino oscillations and the solar neutrino oscillations with the LMA solution of $\sin^2 2\theta_\odot \sim 0.8$.

These results reflect the following general property of our model. The heavy lepton interactions with the Higgs scalar of $\chi$ conserve $S_{2L}$ even after $\chi$ develops a VEV. Therefore, the heavy lepton contributions enhance the $S_{2L}$-conserved property of $\sigma=-1$ and $U_{e3}=0$. The appearance of the larger values of $\sin\alpha$ indicates that more contributions from charged leptons are imported. Since charged leptons are admixtures of the $S_{2L}$-symmetric and -antisymmetric states, which also affect the heavy lepton states, the deviation from $\sigma = -1$ and $U_{e3}=0$ becomes significant. Especially, $\sin^2 2\theta_{atm}$ gets reduced. In the present case, the region for $\sin\alpha \mapgeq 0.1$ is excluded.  The result of $U_{e3}$ can be explained by the simple estimate of Eq.(\ref{Eq:Ue3Order}). Since the tree level contributions to $U_{e3}$ are negligible, $U_{e3}$ arises from the radiative contributions, which amount to $\epsilon \sim \delta m_\nu^{rad} \sim 0.005$ eV (as in FIG.\ref{Fig:NuMasses}).  We find that $U_{e3}(\sim {\rm a~few}\times \epsilon) \sim 0.01-0.02$, which coincides with the result of FIG.\ref{Fig:Ue3} for $\sin\alpha\mapleq 0.1$.
\section{\label{sec:6}Summary}
Summarizing our discussions, we have constructed an $SU(3)_L \times U(1)_N$ gauge model with the lepton triplets of ($\nu^i$, $\ell^i$, $\kappa^i$) for $i=e,\mu,\tau$, where $\kappa^i$ represents positively charged heavy leptons, that provides the LMA solution for the solar neutrino problem compatible with $\sin^2 2\theta_\odot \sim 0.8$. The neutrino masses arise from the tree level neutrino masses induced by the type II seesaw mechanism based on the interactions of lepton triplets with an $SU(3)$-sextet scalar and from the one-loop level masses induced by the Zee type mechanism. The almost maximal atmospheric neutrino mixing and the large solar neutrino mixing are naturally explained by the following mechanisms:

\begin{enumerate}
\item The atmospheric neutrino mixing controlled by the tree level masses and by the radiatively induced masses turns out to be almost maximal because of the presence of an $S_{2L}$ permutation symmetry for left-handed $\mu$ and $\tau$ families with a $Z_4$ discrete symmetry. The bimaximal structure is caused by almost degeneracy between $\delta m_\nu^{e\mu}$ and $\delta m_\nu^{e\tau}$. This degeneracy is assured to occur as a result of our dynamics of the heavy lepton interactions respecting $S_{2L}$.  The suppressed charged lepton contributions are characterized by $\delta m_\nu^{Ce\mu}/\delta m_\nu^{Ce\tau}\sim m_\mu/m_\tau$.

\item The large solar neutrino mixing is caused by generating the almost same contributions of the radiatively induced neutrino masses, which just call for the approximate equality of $f_{\mu\tau}=\mathcal{O}(f_{e\ell})$. 
\end{enumerate}

By performing numerical calculations, we have found that our model has solutions of $\Delta m_{atm}^2 \sim 3.0 \times 10^{-3}$ eV$^2$ with the nearly maximal mixing with $\sin^2 2 \theta_{atm} \mapgeq 0.9$ for atmospheric neutrino oscillations and $\Delta m_\odot^2 \sim 2.0-4.0 \times 10^{-5} $ eV$^2$ with $\sin^2 2\theta_\odot = 0.67-0.98$, corresponding to $\tan^2\theta_\odot=0.27-0.75$, for solar neutrino oscillations. The presence of the charged lepton mixing angle $\alpha$ leads to the deviation of $\sin^2 2\theta_{atm}=1.0$ and to the appearance of the non-vanishing $U_{e3}$. We have estimated $\sin^2 2\theta_{atm}$ and $U_{e3}$ in the three cases of $\sin \alpha=0.0,0.1,0.2$. It turns out that $\sin^2 2\theta_{atm} \mapgeq 0.9$ is obtained for $\sin \alpha \mapleq 0.1$ and $U_{e3}$ is kept sufficiently smaller than the experimental upper bound. 

Finally, we note again that the Yukawa interactions of the heavy leptons with $\chi$ conserve $S_{2L}$ even after the spontaneous breaking. Therefore, if there are only contributions from heavy leptons, the appearance of the maximal atmospheric neutrino mixing is guaranteed by the direct reflection of the $S_{2L}$-conserved mass term of $\nu_e(\nu_\mu+\nu_\tau)$. The charged-lepton interactions, which spoil the $S_{2L}$-conservation, induce their deviations characterized by $\sin \alpha \neq 0$, whose ranges are constrained by $\sin \alpha \mapleq 0.1$. Since the essence in our discussions lies in the presence of the $S_{2L}$ permutation symmetry, we hope that one of the characteristic features behind in the neutrino oscillations is the appearance of this symmetry.

\appendix


\noindent

\begin{center}
\textbf{Table Captions}
\end{center}
\begin{description}
\item{TABLE \ref{Tab:particlesAndSymmetries}:} 
Particle contents with $S_{L2}$, $Z_4$, $L$ and $L_e$ quantum numbers for leptons and Higgs scalars, where $S_{2L}=+(-)$ denotes symmetric (antisymmetric) states. 

\item{TABLE \ref{Tab:S2Z4LLeforYukawa}:} 
$S_{2L}$, $Z_4$, $L$ and $L_e$ quantum numbers of the possible Yukawa interactions. 

\item{TABLE \ref{Tab:S2Z4LLeforHiggsInteractions}:} 
$S_{2L}$, $Z_4$, $L$ and $L_e$ quantum numbers for Higgs interaction relevant for the radiatively induced neutrino masses (except for $\eta \eta s^c s^c$).

\item{TABLE \ref{Tab:FixedParameters}:}
Fixed model parameters, where masses are given in the unit of $v_{weak}=(2\sqrt{2}G_F)^{-1/2}=174$ GeV and $e$ stands for the electromagnetic coupling.   
\end{description}

\noindent

\begin{center}
\textbf{Figure Captions}
\end{center}
\begin{description}
\item{FIG.\ref{Fig:DivergentTwoLoop}:}
Divergent two-loop diagram for $\nu_L^e$-$\nu_L^e$.
\item{FIG.\ref{Fig:ChargedOneLoops}:}
Charged lepton mediated one-loop diagrams.
\item{FIG.\ref{Fig:HeavyOneLoops}:}
Heavy lepton mediated one-loop diagrams. 
\item{FIG.\ref{Fig:SuppressedLoops}:}
Suppressed one- and two-loop diagrams. 	
\item{FIG.\ref{Fig:Sin}:}
Corrected atmospheric and solar neutrino mixing angles $\sin^22\theta_{atm}(\equiv \sin^22\theta_{12}^{new})$ with $\theta_{23}^{new}=\theta_{23}+\xi_{23}$ for ``atm" and similary for $\sin^22\theta_\odot(\equiv \sin^22\theta_{12}^{new})$ for ``solar".
\item{FIG.\ref{Fig:NuMassesC_H}:}
Charged and heavy lepton contributions to the radiative masses for $\nu_e$-$\nu_\mu$  and $\nu_e$-$\nu_\tau$.
\item{FIG.\ref{Fig:HeavyMasses}:}
Heavy lepton masses.
\item{FIG.\ref{Fig:NuMasses}:}
Tree level and radiatively induced neutrino masses.
\item{FIG.\ref{Fig:Xi}:}
Corrections of $\xi_{12,23}$ compared with the mixing angles of $\theta_{12,23}$.
\item{FIG.\ref{Fig:Ue3C_H}:}
Same as in FIG.\ref{Fig:NuMassesC_H} but for the radiative masses for $\nu_e$-$\nu_\mu$  and $\nu_e$-$\nu_\tau$ measured in $U_{e3}$.
\item{FIG.\ref{Fig:Ue3}:}
$\alpha$-dependence of $U_{e3}$. 	
\item{FIG.\ref{Fig:M1m2m3}:}
Neutrino mass eigenvalues. 
\item{FIG.\ref{Fig:DeltaM}:}
Squared mass differences for atmospheric and solar neutrino oscillations. 
\end{description}

\noindent

\begin{center}
\textbf{Table Captions}
\end{center}
\begin{description}
\item{TABLE \ref{Tab:particlesAndSymmetries}:} 
Particle contents with $S_{L2}$, $Z_4$, $L$ and $L_e$ quantum numbers for leptons and Higgs scalars, where $S_{2L}=+(-)$ denotes symmetric (antisymmetric) states. 

\item{TABLE \ref{Tab:S2Z4LLeforYukawa}:} 
$S_{2L}$, $Z_4$, $L$ and $L_e$ quantum numbers of the possible Yukawa interactions. 

\item{TABLE \ref{Tab:S2Z4LLeforHiggsInteractions}:} 
$S_{2L}$, $Z_4$, $L$ and $L_e$ quantum numbers for Higgs interaction relevant for the radiatively induced neutrino masses (except for $\eta \eta s^c s^c$).

\item{TABLE \ref{Tab:FixedParameters}:}
Various parameters used to derive neutrino mixings, where masses ($v$'s, $m$'s and $\mu$'s) are given in the unit of $v_{weak}=(2\sqrt{2}G_F)^{-1/2}=174$ GeV and $e$ stands for the electromagnetic coupling.
\end{description}

\noindent

\begin{center}
\textbf{Figure Captions}
\end{center}
\begin{description}
\item{FIG.\ref{Fig:DivergentTwoLoop}:}
Divergent two-loop diagram for $\nu_L^e$-$\nu_L^e$.
\item{FIG.\ref{Fig:ChargedOneLoops}:}
Charged lepton mediated one-loop diagrams.
\item{FIG.\ref{Fig:HeavyOneLoops}:}
Heavy lepton mediated one-loop diagrams. 
\item{FIG.\ref{Fig:SuppressedLoops}:}
Suppressed one- and two-loop diagrams. 	
\item{FIG.\ref{Fig:Sin}:}
Corrected atmospheric and solar neutrino mixing angles $\sin^22\theta_{atm}(\equiv \sin^22\theta_{12}^{new})$ with $\theta_{23}^{new}=\theta_{23}+\xi_{23}$ for ``atm" and similary for $\sin^22\theta_\odot(\equiv \sin^22\theta_{12}^{new})$ for ``solar".
\item{FIG.\ref{Fig:NuMassesC_H}:}
Charged and heavy lepton contributions to the radiative masses for $\nu_e$-$\nu_\mu$  and $\nu_e$-$\nu_\tau$.
\item{FIG.\ref{Fig:HeavyMasses}:}
Heavy lepton masses.
\item{FIG.\ref{Fig:NuMasses}:}
Tree level and radiatively induced neutrino masses.
\item{FIG.\ref{Fig:Xi}:}
Corrections of $\xi_{12,23}$ compared with the mixing angles of $\theta_{12,23}$.
\item{FIG.\ref{Fig:Ue3C_H}:}
Same as in FIG.\ref{Fig:NuMassesC_H} but for the radiative masses for $\nu_e$-$\nu_\mu$  and $\nu_e$-$\nu_\tau$ measured in $U_{e3}$.
\item{FIG.\ref{Fig:Ue3}:}
$\alpha$-dependence of $U_{e3}$. 	
\item{FIG.\ref{Fig:M1m2m3}:}
Neutrino mass eigenvalues. 
\item{FIG.\ref{Fig:DeltaM}:}
Squared mass differences for atmospheric and solar neutrino oscillations. 
\end{description}

\begin{table}[!htbp]
    \caption{Particle contents with $S_{2L}$, $Z_4$, $L$ and $L_e$ quantum numbers for leptons and Higgs scalars, where $S_{2L}=+(-)$ denotes symmetric (antisymmetric) states.
	\label{Tab:particlesAndSymmetries}}
    \begin{center}
    \begin{tabular}{cccccccccccccc}
    \hline
            & $\psi_L^e, e_R$  & $\psi_L^+$ & $\psi_L^-$ &$\mu_R, \tau_R$ & $\kappa_R^e$ & $\kappa_R^+$ & 
			  $\kappa_R^-$ &$\eta$ & $\eta^\prime$ & $\rho$ &$\rho^\prime$ & $\chi$ & $s$ \\
    \hline
	    $S_{2L}$   & $+$  & $+$  & $-$  &$+$ & $+$  & $+$  & $-$  & $+$ &    $-$  &$+$ & $-$& $+$  & $+$    \\      
	    $Z_4$      & $-i$ & $1$  & $i$  & $1$& $1$ & $i$ & $-1$  & $i$ &    $-i$ &$1$   & $i$& $-i$ & $-1$  \\ 
    \hline
		$L$        & $1$  & $1$  & $1$  &$1$ & $1$  & $1$  & $1$  & $0$ &    $-2$ &$0$  & $0$& $0$  & $-2$   \\  
	    $2L_e$     & $2$  & $0$  & $0$  &$0$ & $2$  & $0$  & $0$  & $0$ &    $0$  &$0$  & $0$& $0$  & $0$    \\    
    \hline
    \end{tabular}
    \end{center}
\end{table}

\begin{table}[!htbp]
    \caption{$S_{2L}$, $Z_4$, $L$ and $L_e$ quantum numbers of the possible Yukawa interactions.
	\label{Tab:S2Z4LLeforYukawa}}
    \begin{center}
    \begin{tabular}{cccccccccc}
    \hline
            & $\overline{(\psi_L^e)^c}\eta \psi_L^+$        & $\overline{(\psi_L^e)^c}\eta \psi_L^-$ 
		    & $\overline{(\psi_L^+)^c}\eta \psi_L^-$        & $\overline{(\psi_L^e)^c}\eta^\prime \psi_L^+$
			& $\overline{(\psi_L^e)^c}\eta^\prime \psi_L^-$ & $\overline{(\psi_L^+)^c}\eta^\prime \psi_L^-$\\
    \hline  
	    $S_{2L}$   & $+$  & $-$ & $-$  & $+$  & $+$  & $+$ \\    
	    $Z_4$      & $1$  & $i$ & $-1$ & $-1$ & $-i$ & $1$ \\ 
    \hline
	    $L$        & $2$  & $2$  & $2$ & $0$  & $0$  & $0$ \\    
	    $2L_e$     & $2$  & $2$  & $0$ & $2$  & $2$  & $0$ \\   
    \hline
    \hline
            & $\overline{\psi_L^e}\rho e_R$    & $\overline{\psi_L^e}\rho \mu_R$ 
			& $\overline{\psi_L^e}\rho \tau_R$ & $\overline{\psi_L^+}\rho \mu_R$
			& $\overline{\psi_L^+}\rho \tau_R$ & $\overline{\psi_L^-}\rho \mu_R$
			& $\overline{\psi_L^-}\rho \tau_R$ \\
    \hline   
	    $S_{2L}$ & $+$ & $+$  & $+$  & $+$ & $+$ & $-$  & $-$ \\ 
	    $Z_4$    & $1$ & $i$  & $i$  & $1$ & $1$ & $-i$ & $-i$ \\ 
	\hline
	    $L$      & $0$ & $0$  & $0$  & $0$ & $0$ & $0$  & $0$ \\   
	    $2L_e$   & $0$ & $-2$ & $-2$ & $0$ & $0$ & $0$  & $0$ \\   
    \hline
    \hline
            & $\overline{\psi_L^e}\rho^\prime e_R$    & $\overline{\psi_L^e}\rho^\prime \mu_R$ 
			& $\overline{\psi_L^e}\rho^\prime \tau_R$ & $\overline{\psi_L^+}\rho^\prime \mu_R$
			& $\overline{\psi_L^+}\rho^\prime \tau_R$ & $\overline{\psi_L^-}\rho^\prime \mu_R$
			& $\overline{\psi_L^-}\rho^\prime \tau_R$ \\
    \hline   
	    $S_{2L}$ & $-$  & $-$  & $-$  & $-$  & $-$ & $+$ & $+$\\ 
	    $Z_4$    & $i$  & $-1$ & $-1$ & $i$  & $i$ & $1$ & $1$ \\ 
	\hline
	    $L$      & $0$  & $0$  & $0$  & $0$  & $0$ & $0$ & $0$\\   
	    $2L_e$   & $0$  & $-2$ & $-2$ & $0$  & $0$ & $0$ & $0$\\   
    \hline
    \hline
            & $\overline{\psi_L^e}\chi \kappa_R^e$ & $\overline{\psi_L^e}\chi \kappa_R^+$ 
			& $\overline{\psi_L^e}\chi \kappa_R^-$ & $\overline{\psi_L^+}\chi \kappa_R^+$
			& $\overline{\psi_L^-}\chi \kappa_R^-$ & $\overline{\psi_L^+}\chi \kappa_R^-$
			& $\overline{\psi_L^-}\chi \kappa_R^+$  \\
    \hline   
	    $S_{2L}$   & $+$  & $+$  & $-$  & $+$  & $+$ & $-$  & $-$ \\ 
	    $Z_4$      & $1$  & $i$ & $-1$ & $1$  & $1$ & $i$ & $-i$ \\ 
	\hline
	    $L$        & $0$  & $0$  & $0$  & $0$  & $0$ & $0$  & $0$ \\   
	    $2L_e$     & $0$  & $-2$ & $-2$ & $0$  & $0$ & $0$  & $0$ \\   
    \hline
    \hline
            & $\overline{(\psi_L^e)^c}s \psi_L^e$ & $\overline{(\psi_L^e)^c}s \psi_L^+$ 
			& $\overline{(\psi_L^e)^c}s \psi_L^-$ & $\overline{(\psi_L^+)^c}s \psi_L^+$
			& $\overline{(\psi_L^-)^c}s \psi_L^-$ & $\overline{(\psi_L^+)^c}s \psi_L^-$\\
    \hline  
	    $S_{2L}$   & $+$  & $+$  & $-$  & $+$  & $+$ & $-$  \\    
	    $Z_4$      & $1$  & $i$  & $-1$ & $-1$ & $1$ & $1$ \\ 
    \hline
	    $L$        & $0$  & $0$  & $0$  & $0$  & $0$ & $0$ \\    
	    $2L_e$     & $4$  & $2$  & $2$  & $0$  & $0$ & $0$  \\   
    \hline 
    \end{tabular}
    \end{center}
\end{table}

\begin{table}[!htbp]
    \caption{$S_{2L}$, $Z_4$, $L$ and $L_e$ quantum numbers for Higgs interaction relevant for the radiatve induced neutrino masses (except $\eta \eta s^c s^c$).   
	\label{Tab:S2Z4LLeforHiggsInteractions}}
    \begin{center}
    \begin{tabular}{cccccccc}
    \hline
            & $\eta s \eta$
            & $\eta^\prime s \eta^\prime$
            & $(\eta^\dagger \rho^\prime)(\eta^{\prime\dagger} \chi)$
            & $(\eta^{\prime\dagger} \rho^\prime)(\eta^\dagger \chi)$
            & $(\chi^\dagger\eta)(\eta^\dagger\chi)$
            & $(\eta^{\prime\dagger}\chi)(\chi^\dagger\eta^\prime)$
			& $\eta \eta s^c s^c$ \\
    \hline
	    $S_{2L}$ & $+$  & $+$ & $+$ & $+$ & $+$ & $+$ & $+$  \\      
	    $Z_4$    & $1$  & $1$ & $1$ & $1$ & $1$ & $1$ & $-1$ \\ 
	\hline
	    $L$      & $-2$ & $-6$ & $2$ & $2$ & $0$ & $0$ & $0$\\    
	    $2L_e$   & $0$  & $0$  & $0$ & $0$ & $0$ & $0$ & $0$\\   
    \hline
    \end{tabular}
    \end{center}
\end{table}

\begin{table}[!htbp]
    \caption{Various parameters used to derive neutrino mixings, where masses ($v$'s, $m$'s and $\mu$'s) are given in the unit of $v_{weak}=(2\sqrt{2}G_F)^{-1/2}=174$ GeV and $e$ stands for the electromagnetic coupling.   
	\label{Tab:FixedParameters}}
    \begin{center}
    \begin{tabular}{ccc|ccc|cccc}
    \hline
              $v_{\eta,\eta^\prime}$
            & $v_{\rho,\rho^\prime}$
            & $v_\chi$
            & $m_{\eta,\eta^\prime,\rho,\rho^\prime}$
            & $m_\chi$
            & $m_s,\mu_{1,2}$
		    & $g_s$
		    & $f_{e\ell}$
		    & $f_{\mu\tau}$
		    & $\lambda_{3,4}$ \\
    \hline
	    $1/20$ & $1/\sqrt{2}$  & $10$ & $1$ & $10$ & $4.6 \times 10^9$ &
	    $e$ & $1.0 \times 10^{-8}$ & $0.5 \times 10^{-8}$ & $0.3$ \\
	\hline
    \end{tabular}
    \end{center}
\end{table}

\newpage
\noindent

\begin{figure}[!htbp]
\begin{flushleft}
 \includegraphics*[3mm,1mm][200mm,130mm]{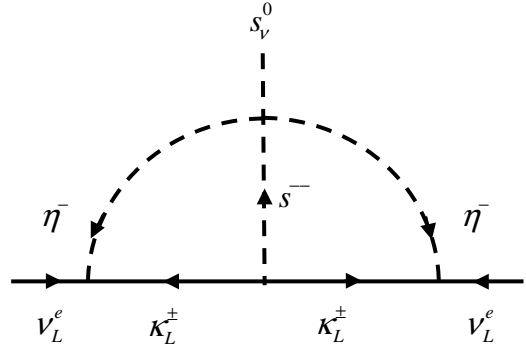}
\end{flushleft}
\caption{Divergent two-loop diagram for $\nu_L^e$-$\nu_L^e$.}
\label{Fig:DivergentTwoLoop}
\end{figure}

\begin{figure}[!htbp]
 \begin{flushleft}
 \includegraphics*[3mm,1mm][200mm,140mm]{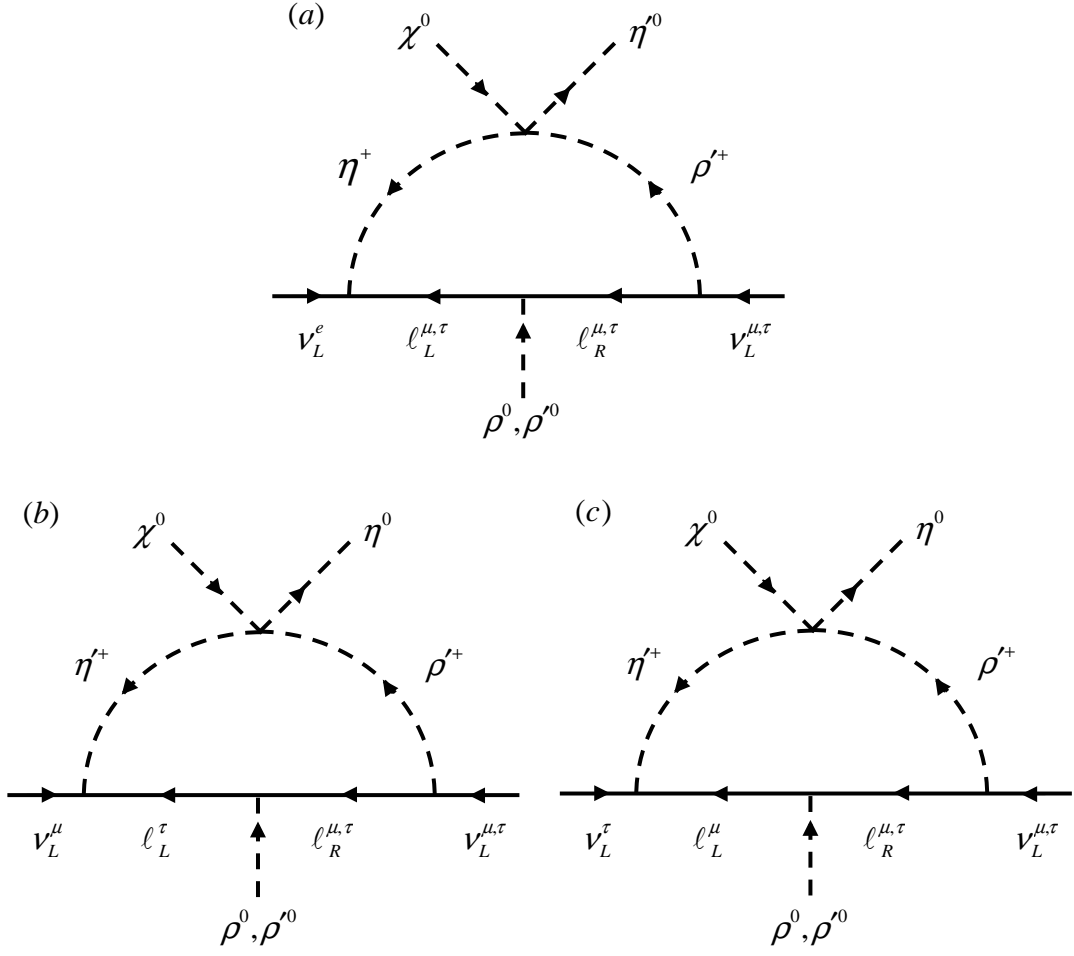}
 \end{flushleft}
  \caption{Charged lepton mediated one-loop diagrams.}
\label{Fig:ChargedOneLoops}
\end{figure}

\begin{figure}[!htbp]
 \begin{flushleft}
 \includegraphics*[3mm,1mm][200mm,140mm]{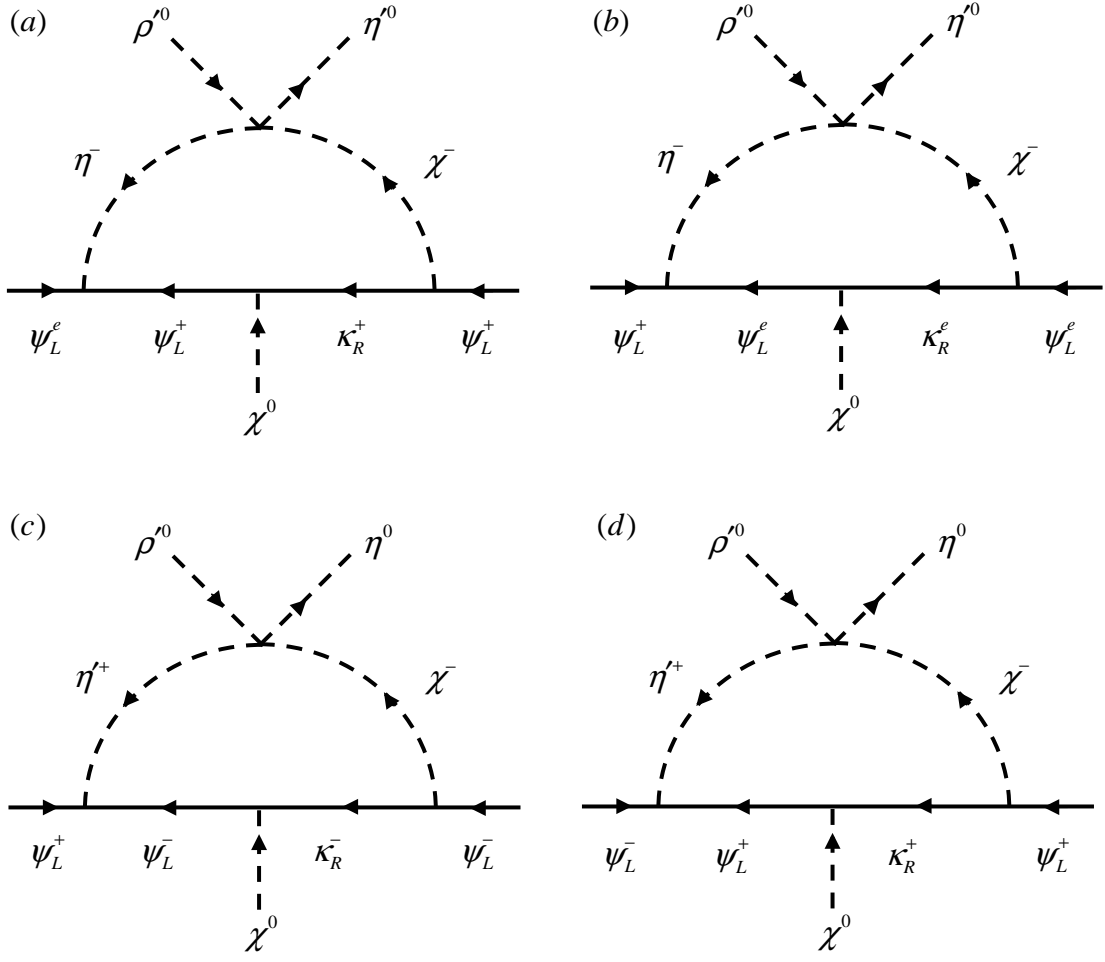}
 \end{flushleft}
  \caption{Heavy lepton mediated one-loop diagrams.}
\label{Fig:HeavyOneLoops}
\end{figure}

\begin{figure}[!htbp]
 \begin{flushleft}
\includegraphics*[3mm,1mm][200mm,130mm]{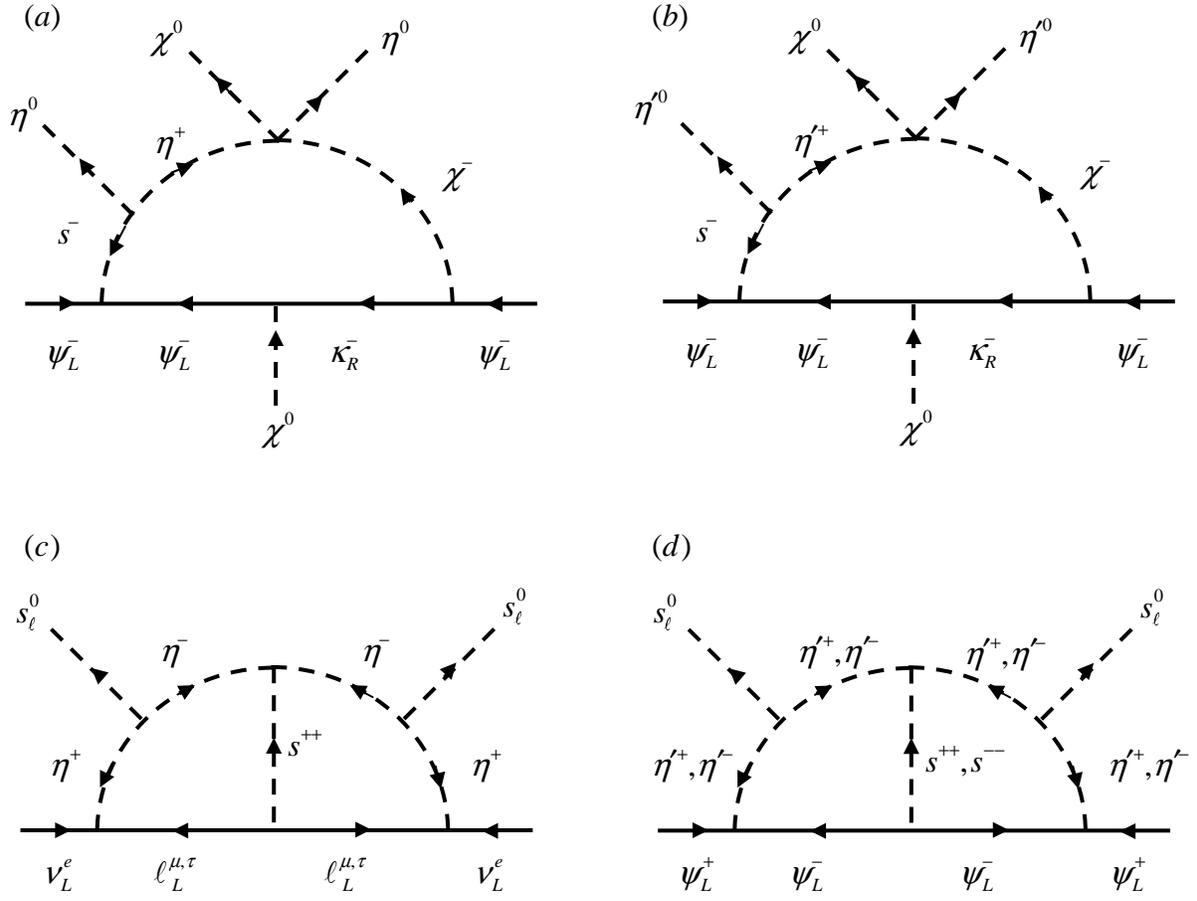}
 \end{flushleft}
  \caption{Suppressed one- and two-loop diagrams}
\label{Fig:SuppressedLoops}
\end{figure}

\begin{figure}[!htbp]
 \begin{flushleft}
\includegraphics*[3mm,13mm][200mm,150mm]{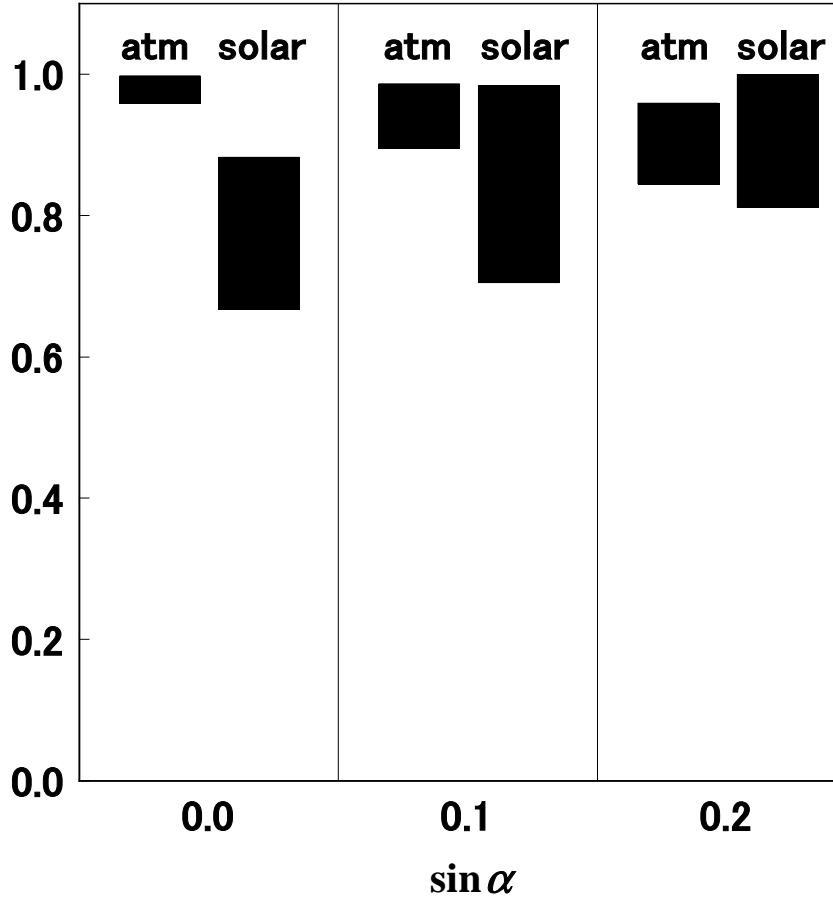}
  \end{flushleft}
  \caption{Corrected atmospheric and solar neutrino mixing angles $\sin^22\theta_{atm}(\equiv \sin^22\theta_{12}^{new})$ with $\theta_{23}^{new}=\theta_{23}+\xi_{23}$ for ``atm" and similary for $\sin^22\theta_\odot(\equiv \sin^22\theta_{12}^{new})$ for ``solar".}\label{Fig:Sin}
\end{figure}

\begin{figure}[!htbp]
  \begin{flushleft}
\includegraphics*[3mm,13mm][200mm,140mm]{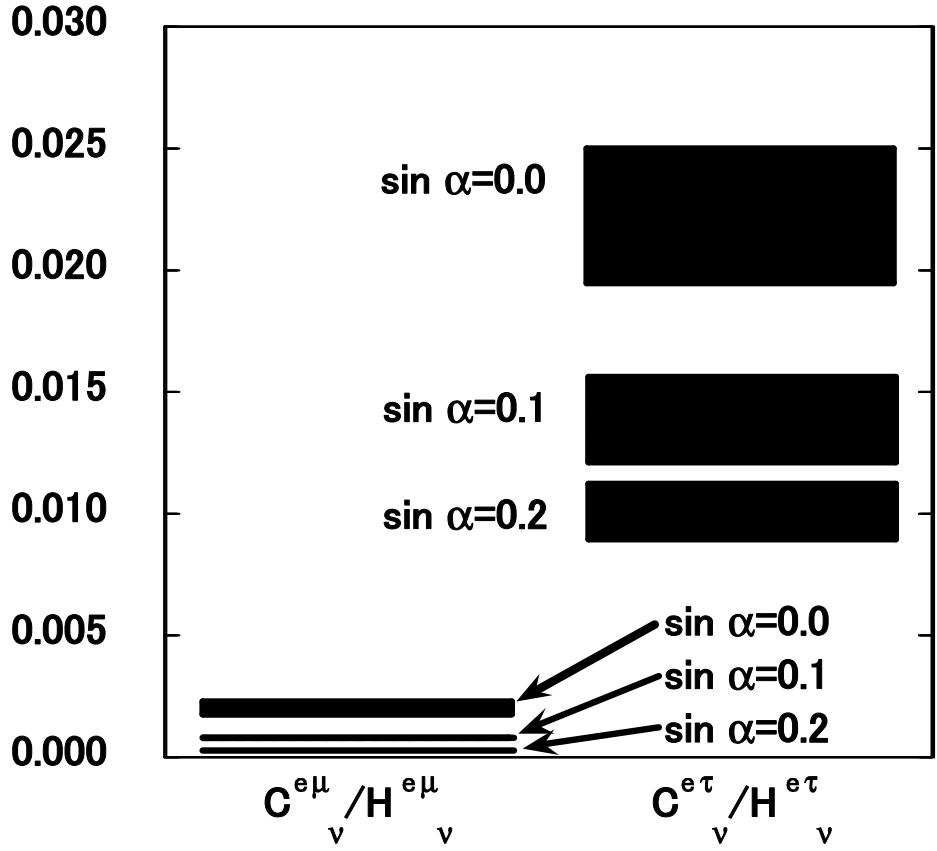}
  \end{flushleft}
  \caption{Charged and heavy lepton contributions to the radiative masses for $\nu_e$-$\nu_\mu$ and $\nu_e$-$\nu_\tau$.}
\label{Fig:NuMassesC_H}
\end{figure}

\begin{figure}[!htbp]
 \begin{flushleft}
\includegraphics*[3mm,13mm][200mm,140mm]{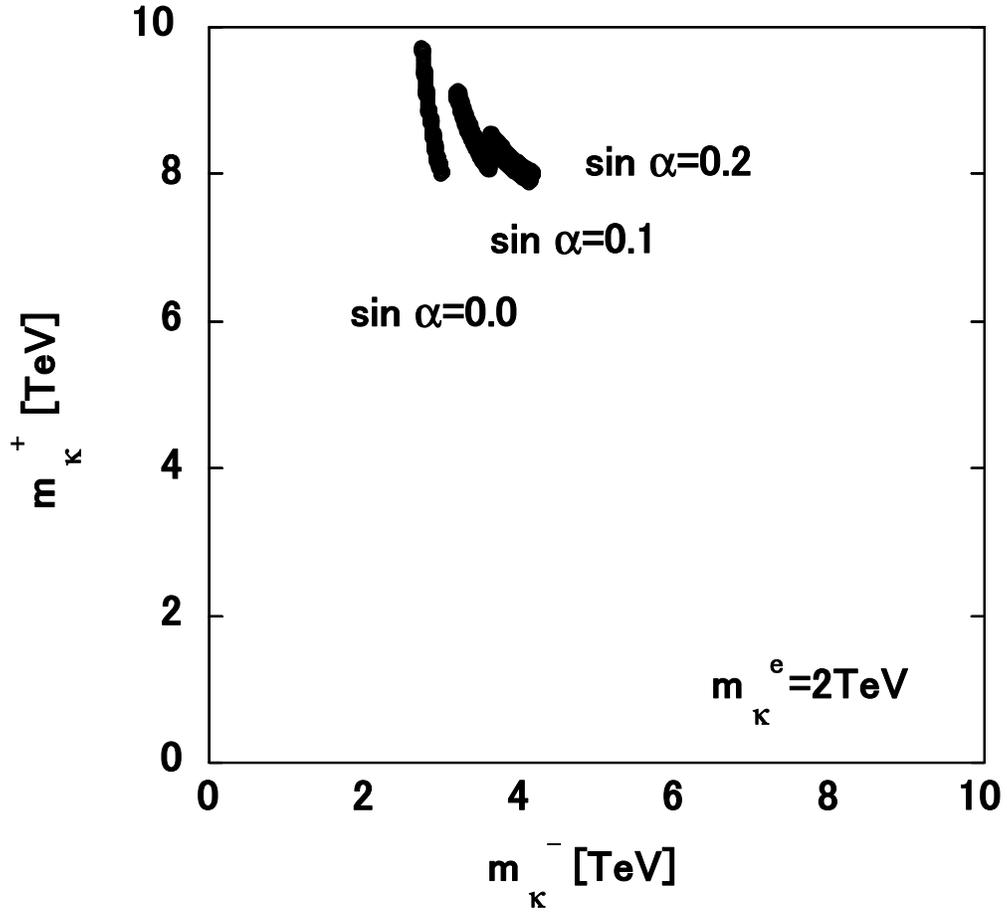}
 \end{flushleft}
  \caption{Heavy lepton masses.}
\label{Fig:HeavyMasses}
\end{figure}

\begin{figure}[!htbp]
 \begin{flushleft}
\includegraphics*[3mm,13mm][200mm,140mm]{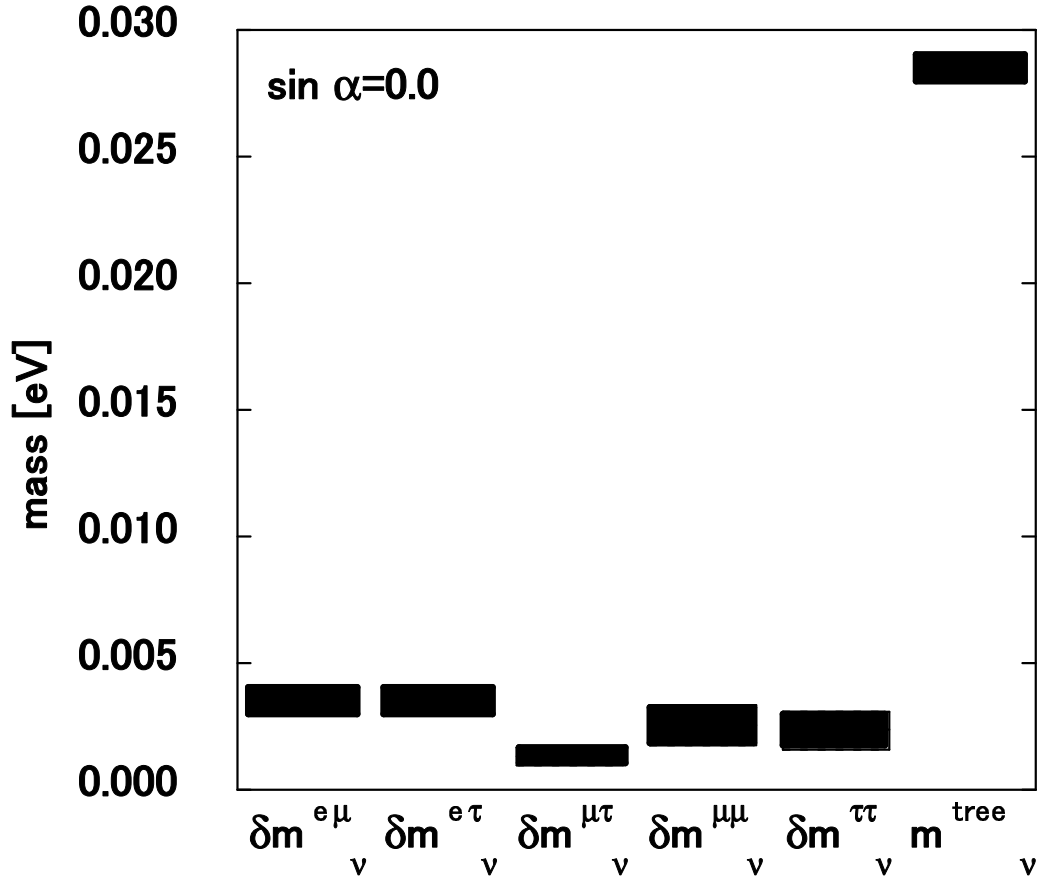}
 \end{flushleft}
  \caption{Tree level and radiatively induced neutrino masses.}
\label{Fig:NuMasses}
\end{figure}

\begin{figure}[!htbp]
  \begin{flushleft}

\includegraphics*[3mm,10mm][160mm,150mm]{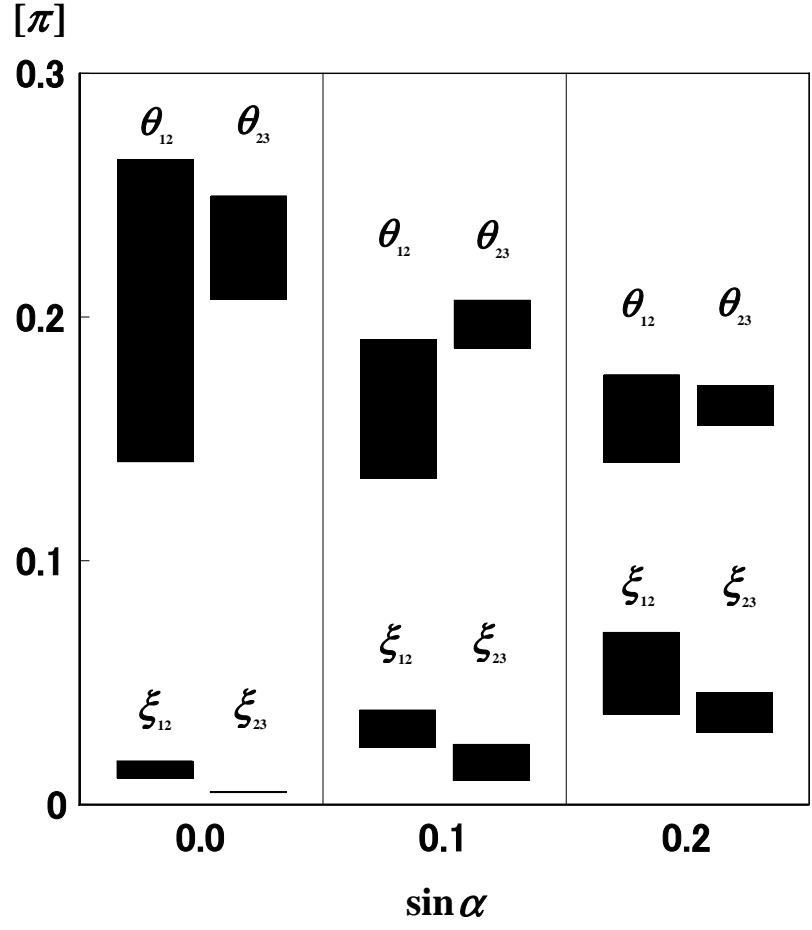}
  \end{flushleft}
  \caption{Corrections of $\xi_{12,23}$ compared with the mixing angles of $\theta_{12,23}$ in the unit of $\pi$.}
\label{Fig:Xi}
\end{figure}

\begin{figure}[!htbp]
  \begin{flushleft}
\includegraphics*[3mm,13mm][200mm,130mm]{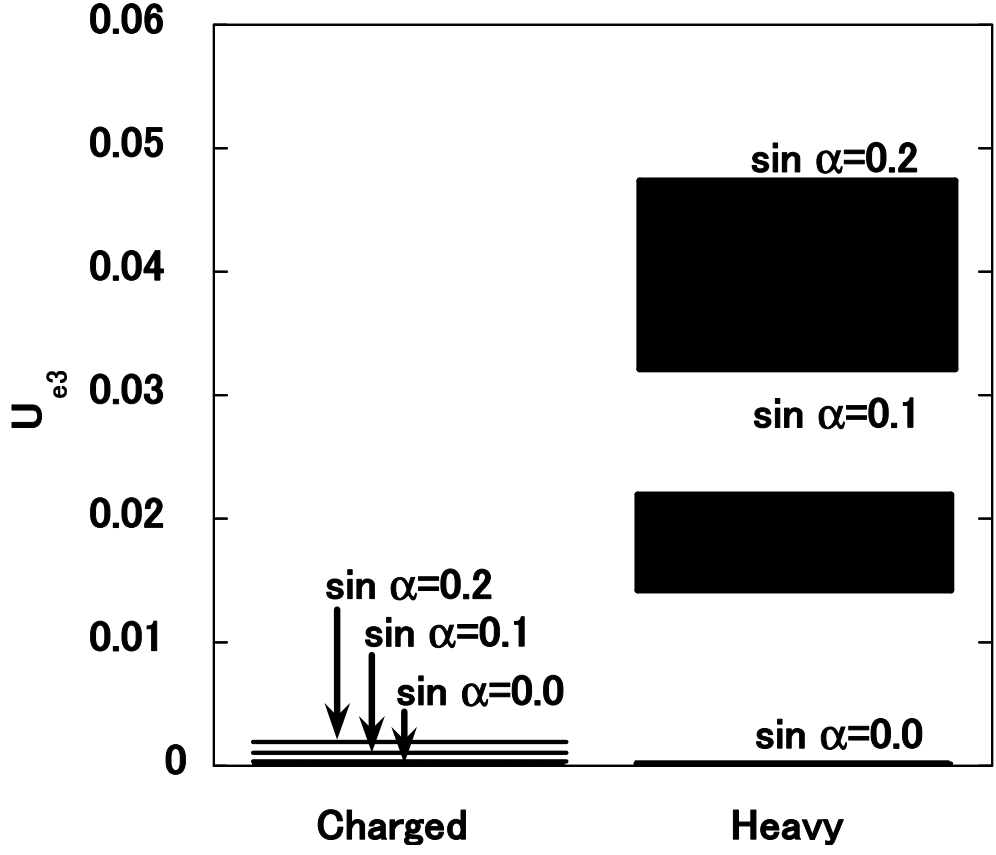}
  \end{flushleft}
  \caption{Same as in FIG.\ref{Fig:NuMassesC_H} but for the radiative masses for $\nu_e$-$\nu_\mu$  and $\nu_e$-$\nu_\tau$s measured in $U_{e3}$.}
\label{Fig:Ue3C_H}
\end{figure}

\begin{figure}[!htbp]
  \begin{flushleft}
\includegraphics*[3mm,13mm][200mm,140mm]{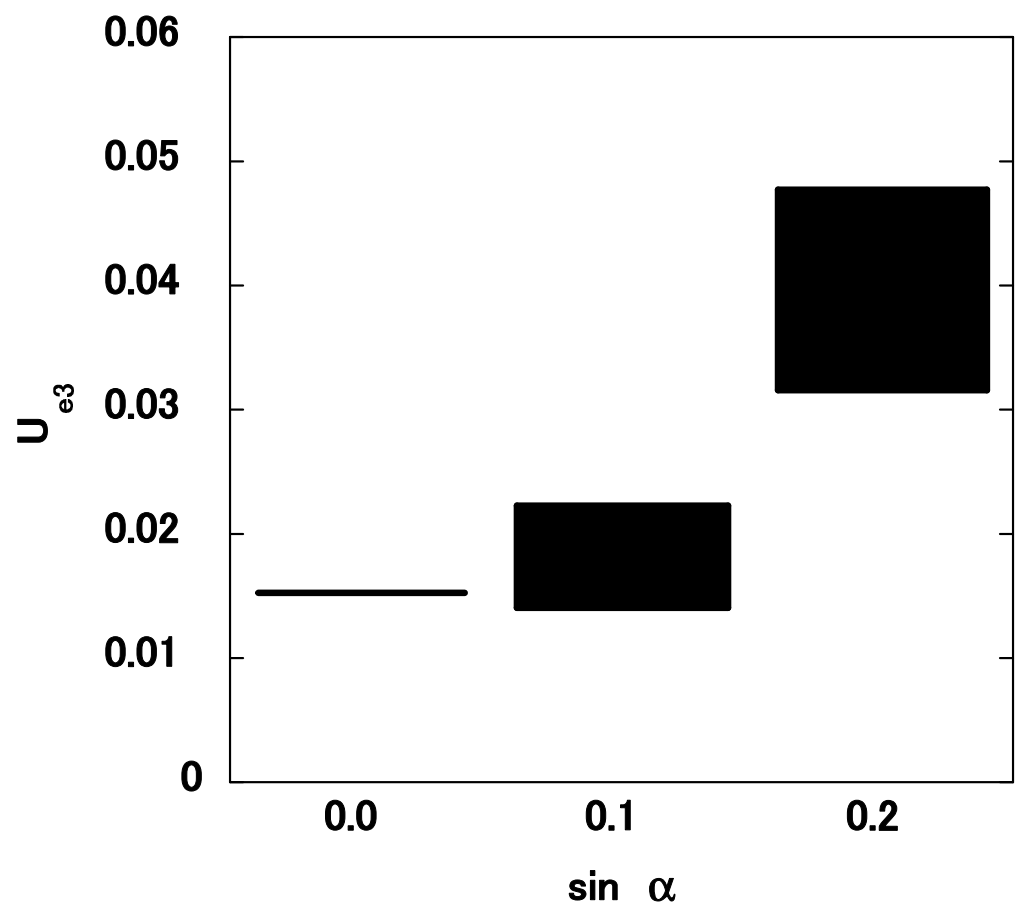}
  \end{flushleft}
  \caption{$\alpha$-dependence of $U_{e3}$.}
\label{Fig:Ue3}
\end{figure}

\begin{figure}[!htbp]
  \begin{flushleft}
\includegraphics*[3mm,10mm][200mm,140mm]{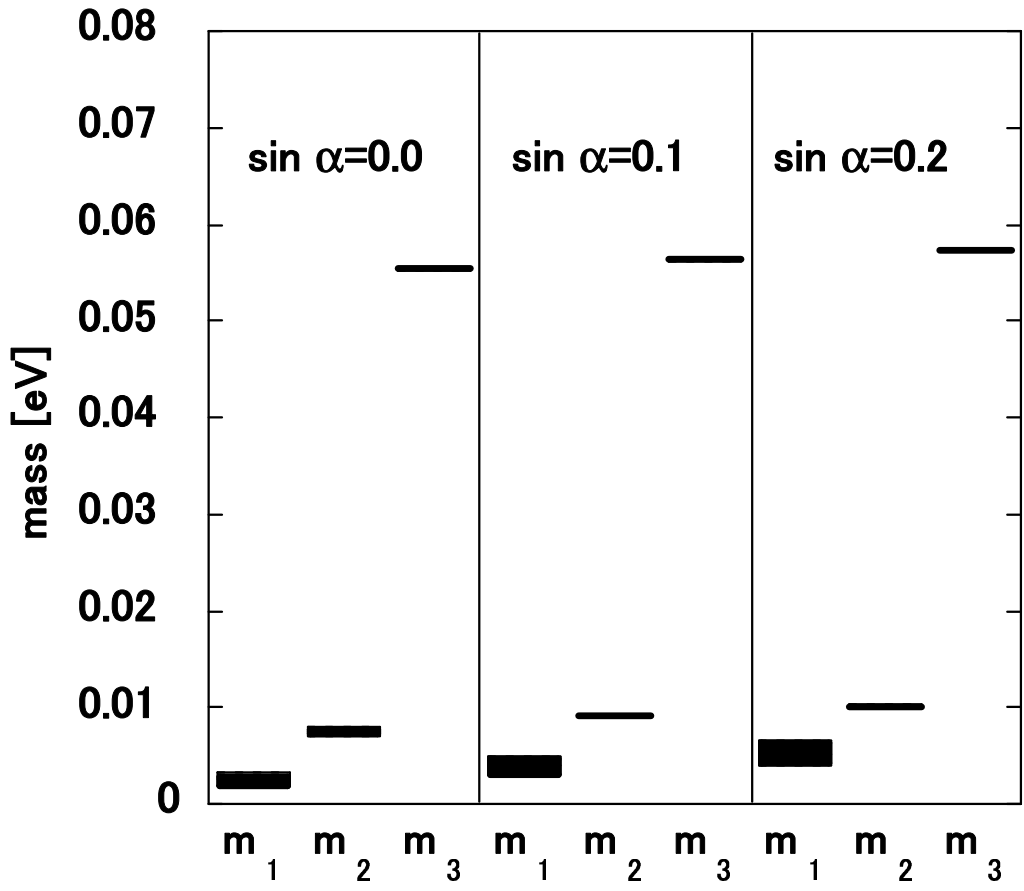}
 \end{flushleft}
  \caption{Neutrino mass eigenvalues.}
\label{Fig:M1m2m3}
\end{figure}

\begin{figure}[!htbp]
  \begin{flushleft}
\includegraphics*[3mm,10mm][200mm,150mm]{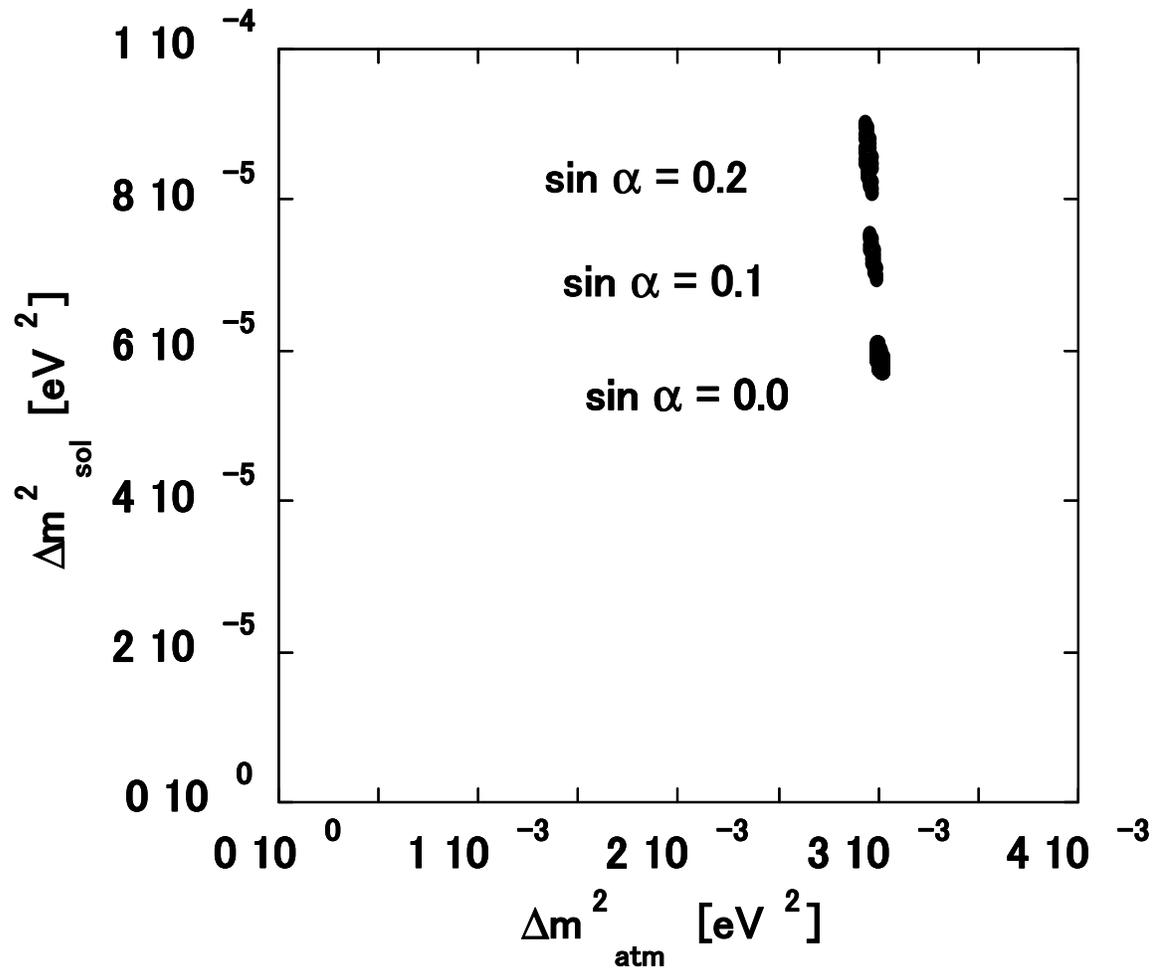}
  \end{flushleft}
  \caption{Squared mass differences for atmospheric and solar neutrino oscillations.}
\label{Fig:DeltaM}
\end{figure}

\end{document}